\documentclass[useAMS]{./aa}
\pdfoutput=1
\usepackage[varg]{txfonts}
\usepackage{natbib}
\bibpunct{(}{)}{;}{a}{}{,}
\usepackage{color}
\usepackage{longtable}
\usepackage{multicol}
\usepackage{graphicx}

\newcommand{\simgt}{\lower.5ex\hbox{$\; \buildrel > \over \sim \;$}}
\newcommand{\simlt}{\lower.5ex\hbox{$\; \buildrel < \over \sim \;$}}

\def\farcs{\hbox{$.\!\!^{\prime\prime}$}}

\def\ls{\mathrel{\hbox{\rlap{\hbox{\lower4pt\hbox{$\sim$}}}\hbox{$<$}}}}
\def\gs{\mathrel{\hbox{\rlap{\hbox{\lower4pt\hbox{$\sim$}}}\hbox{$>$}}}}

\begin{document}

\title{Probing the dynamical and X-ray mass proxies of the
  cluster of galaxies Abell S1101\thanks{We have made use of VLT/VIMOS
    observations taken with the ESO Telescope at the Paranal
    Observatory under programme 087.A-0096.}}

\author{Andreas~Rabitz\inst{\ref{inst1}},
        Yu-Ying~Zhang\inst{\ref{inst2}},
        Axel~Schwope\inst{\ref{inst1}},
        Miguel~Verdugo\inst{\ref{inst3}},
        Thomas~H.~Reiprich\inst{\ref{inst2}}, and
        Matthias~Klein\inst{\ref{inst7}}
        }

\institute{Leibniz-Institut f\"ur Astrophysik Potsdam (AIP), An der Sternwarte 16, 14482 Potsdam, Germany\label{inst1}
\and 
Argelander-Institut f\"ur Astronomie, Auf dem H\"ugel 71, 53121 Bonn, Germany\label{inst2}
\and
Department for Astrophysics University of Vienna, T\"urkenschanzstr. 17, 1180 Vienna, Austria\label{inst3}
\and
Max-Planck-Institut f\"ur extraterrestrische Physik, Giessenbachstr. 1, 85748 Garching, Germany\label{inst7}}

\authorrunning{Rabitz et al.}

\titlerunning{Probing the dynamical and X-ray mass proxies of Abell S1101}

\date{Received 02/06/2016 / Accepted 25/08/2016}

\abstract{The galaxy cluster Abell~S1101 (S1101 hereafter) deviates
  significantly from the X-ray luminosity versus velocity dispersion
  relation (\(L-\sigma\)) of galaxy clusters in our previous
  study. Given reliable X-ray luminosity measurement combining
  \emph{XMM-Newton} and \emph{ROSAT}, this could most likely be caused
  by the bias in the velocity dispersion due to interlopers and low
  member statistic in the previous sample of member galaxies, which
  was solely based on 20 galaxy redshifts drawn from the literature.}
{We intend to increase the galaxy member statistic to perform
  a precision measurement of the velocity dispersion and dynamical mass of
  S1101. We aim for a detailed substructure and dynamical state characterization
  of this cluster, and a comparison of mass estimates derived from (i) the
  velocity dispersion (\(M_{vir}\)), (ii) the caustic mass computation
  (\(M_{caustic}\)), and (iii) mass proxies from X-ray observations and the
  Sunyaev-Zel’dovich (SZ) effect.}
  {We carried out new optical spectroscopic observations
  of the galaxies in this cluster field with VIMOS, obtaining a sample of
  \(\sim60\) member galaxies for S1101. We revised the
  cluster redshift and velocity dispersion measurements based on this
  sample and also applied the Dressler-Shectman substructure
  test.}  {The completeness of cluster members within
  \(r_{200}\) was significantly improved for this cluster.
  Tests for dynamical substructure did not show evidence for major
  disturbances or merging activities in S1101. We find good agreement between the dynamical
  cluster mass measurements and X-ray mass estimates which confirms
  the relaxed state of the cluster displayed in the 2D substructure
  test. The SZ mass proxy is slightly higher than the other estimates.
  The updated measurement of \(\sigma\) erased the deviation of S1101
  in the \(L-\sigma\) relation. We also noticed a background structure
  in the cluster field of S1101. It is a galaxy group very close to
  the cluster S1101 in projection but at almost twice its
  redshift. However its mass is too low to significantly bias the
  observed bolometric X-ray luminosity of S1101. Hence, we can
  conclude that the deviation of S1101 in the \(L-\sigma\) relation in
  our previous study can be explained by low member statistic and
  galaxy interlopers, which are known to introduce biases in the
  estimated velocity dispersion.}{} \keywords{Cosmology: observations
  --- Galaxies: clusters: individual: Abell S1101 --- Galaxies:
  clusters: intracluster medium --- Methods: data analysis --- X-rays:
  galaxies: clusters --- Galaxy: kinematics and dynamics}

\maketitle

\section{Introduction}
\label{Introduction}

Galaxy clusters memorize structure formation,
e.g. \citet{Borgani2001}, and cluster surveys have thus been widely
used to constrain the cosmological parameters such as the dark energy
content, e.g. \citet{Mantz2014}. Various methods such as optical,
e.g. \citet{Gladders2005}, X-rays, e.g. \citet{Boehringer2001}, weak lensing,
e.g. \citet{Schneider1996}, and the Sunyaev-Zel'dovich \citep[SZ;][]{Sunyaev1972}
effect, e.g. \citet{Vale2006}, could potentially lead
to biases in determining the mass function in the cluster cosmological
experiments. X-ray selects systems containing hot gas, as a sign of
virialization \citep[e.g.][]{Weinberg2002}, and are considered to
provide a cleaner and more complete selection regarding mass.

To improve our knowledge on the X-ray selection method, we
investigated the X-ray bolometric luminosity versus velocity
dispersion ($L_{\rm bol}-\sigma$) relation of 62 clusters in an X-ray
flux-limited sample of 64 clusters \citep{Zhang2011}.  The observed
scatter in the relation in our observational data was thought to be
mainly related to the presence of cool cores but was even present
after the cool-core correction in deriving the X-ray luminosity.  To
study the other possible origin of this scatter, in \citet{Zhang2011}
we also compared the $L_{\rm bol}-\sigma$ relations between our
observational sample and a sample of 21 clusters and groups from very
high resolution simulations with and without active galactic nuclei
(AGN) feedback provided by \citet{Puchwein2008}. The simulated sample
with AGN feedback matches well with the observational sample of 56
clusters excluding six clusters with less than 45 cluster galaxy
redshifts (\(n_{\rm gal}<45\)), in which AGN feedback in different
phases explains the increasing scatter of our observational sample
toward the low-mass end. We note that the velocity dispersion
estimates for those \(n_{\rm gal}<45\) clusters are about a factor of
two lower than the \(L_{\rm bol}-\sigma\) predictions such that they
cause more than two times larger scatter than that of the remaining
clusters. This indicates that AGN feedback can not be the main origin
of the scatter of the \(L_{\rm bol}-\sigma\) relation for those
\(n_{\rm gal}<45\) systems.  We suspected that systematic
uncertainties in the velocity dispersion estimates (e.g. Biviano et
al. 2006; Saro et al. 2013) may play a stronger role than the cluster
physics such as AGN feedback in causing the scatter. We confirmed our
guess by carefully testing systematic uncertainties using the
simulated sample due to interlopers and the selection of cluster
members regarding the aperture radius and mass limit in
\citet{Zhang2011}. The systematic uncertainties due to those effects
are up to 40\%, in which the low member statistic in recovering the
caustic amplitude may play an important role in causing the bias
\citep[also see][]{Clerc2016sub}.

As a second-order effect, we found in \citet{Zhang2011} that
interlopers always bias the velocity dispersion measurements towards
lower values for our simulated sample except for one cluster. There
are significant fractions of galaxies within 1.2 Abell radii which are
not in the virialized regions for poor systems in our simulated
sample. As also shown in Fig.~10 in \citet{Biviano2006}, unrecognized
interlopers that are outside the virial radius ($r_{\rm vir}$) but
dynamically linked to the host cluster, and that do not form major
substructures, bias the $\sigma$ estimate toward lower values than
cluster galaxies.  There are also similar studies based on other
simulations \citep[e.g.][]{Saro2013}. Particularly, the line-of-sight
velocity of the galaxies with available spectroscopic redshifts
\citep[][]{DeVaucouleurs1991,Shectman1996,Zabludoff1998,Jones2009} as
a function of the projected radius shows a box-shape instead of a
caustic shape, for one of the \(n_{\rm gal}<45\) clusters --- Abell
S1101 (S1101 hereafter). This indicates that the chance of having a
high fraction of unrecognized interlopers at large radii is likely
high for this cluster. To disentangle the scatter driven by
interlopers and by other physics in the \(L_{\rm bol}-\sigma\)
relation and constrain possible bias in the X-ray selection method, we
carried out a detailed study on S1101 based on newly awarded
VLT/VIMOS \citep[Very Large Telescope / VIsible Multi-Object
Spectrograph;][]{LeFevre2003} spectroscopic observations together with
\emph{XMM-Newton} and \emph{ROSAT} X-ray data. We note that S1101 was
detected as RXC J2313.9-4243 in \citet{Boehringer2004} in the
ROSAT All-Sky Survey \citep[RASS;][]{Truemper1992}.
It is also listed as SPT-CL J2313-4243 detected through the SZ effect
in \citet{Bleem2015}.

Throughout the paper we assume a standard cosmology with
\(\Omega_{M}=0.3\), \(\Omega_{\Lambda}=0.7\) and
\(H_{0}=70~\textnormal{km~s}^{-1}~\textnormal{Mpc}^{-1}\).  Thus, with
an angular diameter distance of \(D_{A}=224~\textnormal{Mpc}\) a scale
of \(1\arcmin\) corresponds to an extent of
\(\sim65~\textnormal{kpc}\) at the cluster redshift
\(z_{cl}=0.05601\pm0.00027\).  Errors are given as 95\% confidence
intervals, unless stated otherwise.

\section{Optical imaging and target selection}
\label{imaging}
\begin{figure*}[t]
 \centering \begin{minipage}{8.75cm} \hskip-0.4cm
 \includegraphics[angle=0,width=9cm]{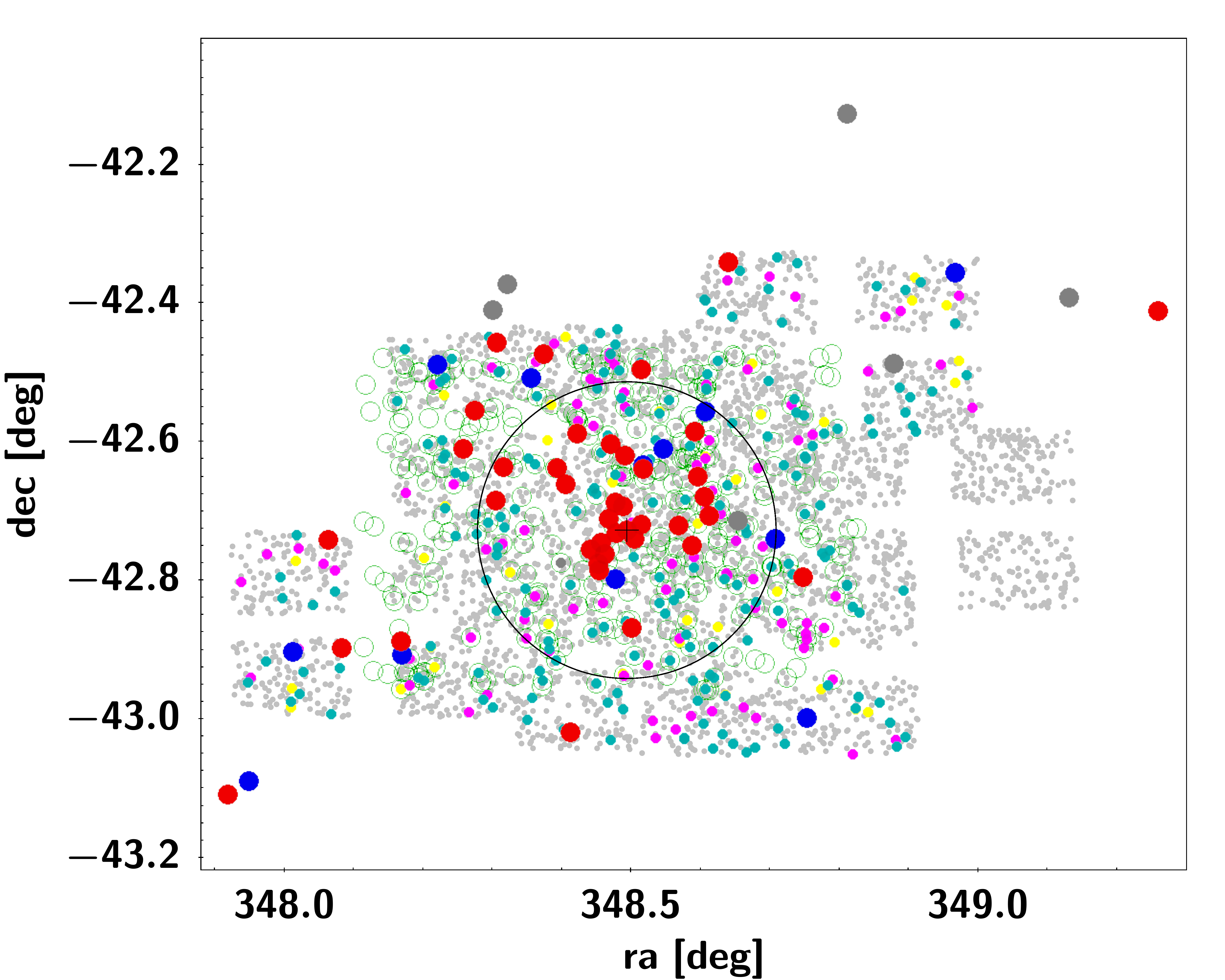}
 \end{minipage}
 \begin{minipage}{8.75cm} \hskip+0.4cm
 \includegraphics[angle=0,width=9cm]{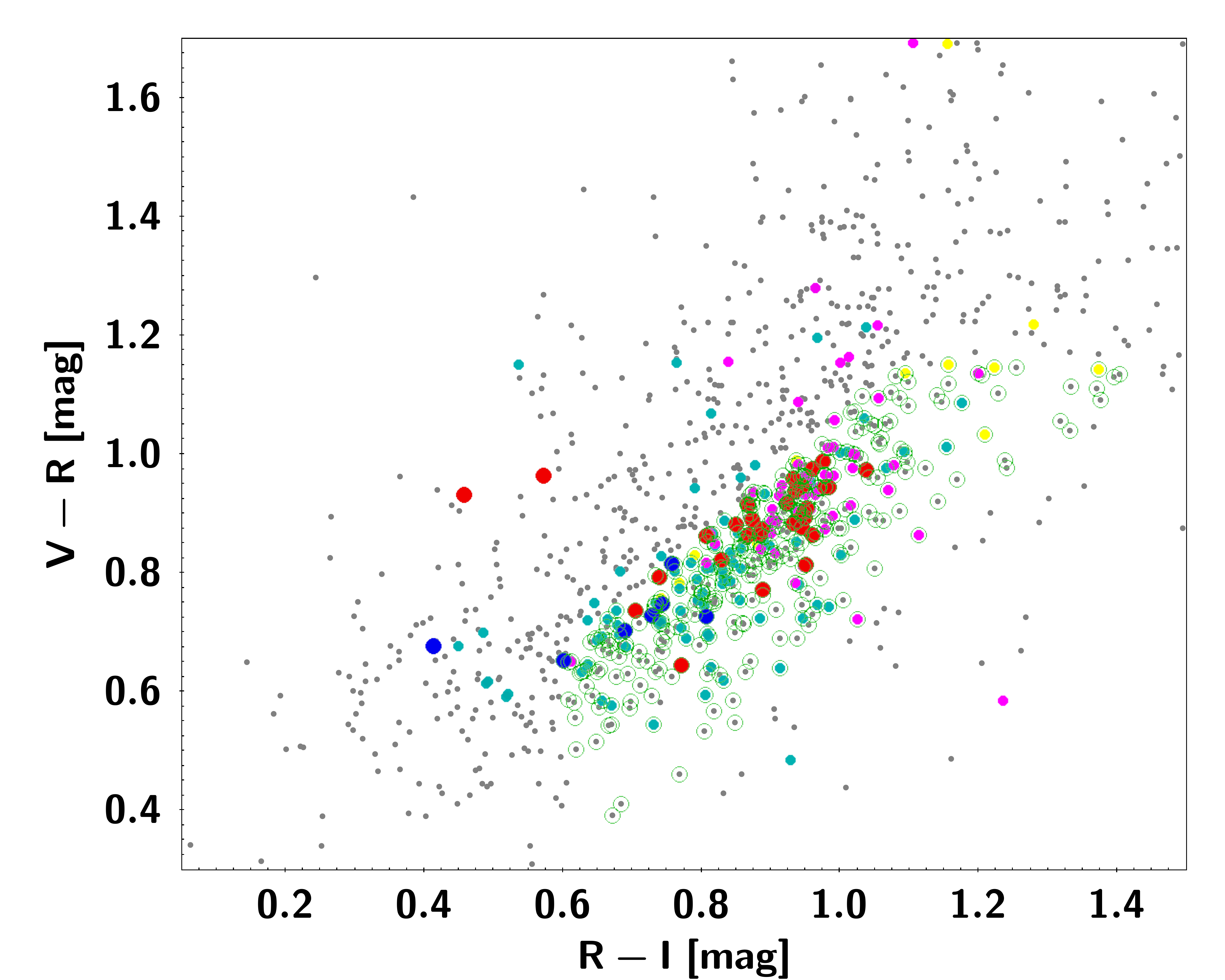}
\end{minipage}
\caption{Skypositions (left panel) and WFI colour-colour diagram (right panel)
  of the objects in the cluster field of S1101. Small gray dots show
  all sources detected from the VIMOS pre-imaging and the WFI-imaging
  for the left and the right panel, respectively. Green circles
  indicate WFI colour-colour selected candidates for the spectroscopic
  follow-up. The imaging candidates that were spectroscopically
  observed but classified as non-members are marked by magenta and
  cyan dots for galaxies with passive and active spectral types and
  yellow dots for stars. Large dots in red, blue and gray correspond
  to spectroscopically verified cluster members with passive, active
  and unknown spectral type. The black cross denotes
  the cluster X-ray centre. The big black circle shows
  the area within a projected distance of $r_{500}$. Due to bad weather,
  the spectroscopic follow-up shows an artificial elongation along the
  southeast-northwest direction as not all scheduled VIMOS pointings in
  the cluster field were performed.}
  \label{specs_pos_type}
\end{figure*}
The cluster S1101 was observed at the ESO/MPG 2.2\,m telescope with
the wide-field imager \citep[WFI; ][]{Baade1999} in the $V/89$-,
$R_{\rm c}/162$, and $I_{\rm c}$-bands under non-photometric
conditions (PI: Reiprich) as bad weather backup target.
We reduced the data using the {\tt THELI} pipeline
\citep{Erben2005,Schirmer2013}. The co-added images have 1\farcs5
seeing in $V/89$ and $R_{\rm c}/162$, and 3\farcs1 seeing in $I_{\rm c}$.
The relatively poor quality of the imaging combined with sub-optimal
weather makes the determination of the zero point not better than 0.15~mag.

The combination of those filters is not the best to bracket the
4000\AA\ break to well isolate the red sequence for the colour
selection of the candidate cluster galaxies at the redshift of
S1101, but was chosen to also select active galaxies.
Furthermore, a precise pre-selection of cluster member galaxies
according to their colours was hindered by the bad seeing and
resulting improper photometric calibration.  The existing 2MASS
\citep[][]{Skrutskie2006} data also failed to improve
the photometric calibration. Note that most 2MASS sources are also
saturated for the galaxies in our observations. The standard fields
are not of sufficient quality to calibrate the B-band
FORS data. 

Still, from the reduced imaging data we are able to extract catalogues
using {\tt{SExtractor}}
\citep[][]{Bertin1996} in dual-image mode, with $R_{\rm c}/162$ band
as detection image. We filtered the catalogue for objects where the
{\tt{CLASS\_STAR}} of {\tt{SExtractor}} indicates a likely galaxy and
the fluxes in the respective bands are less than those of the apparent
bright central galaxy (BCG).

Our VIMOS spectroscopic follow-up covers actually a larger sky area
than the WFI field. Therefore, we have to rely on the VIMOS pre-imaging, 
performed with the \(R\) filter, to select
objects which appear like bright ellipticals for the incremental
region. We reduced the pre-imaging of our VIMOS campaign also with the
{\tt THELI} pipeline. The photometry is calibrated against the
SuperCOSMOS catalogues in the Vega system. The offset
between SuperCOSMOS and the calibrated VIMOS \(R\) band catalogue is
about 0.3~mag, which is rather significant. It may be caused by
calibrations in either of the observations and/or filter differences.
Nevertheless, we are able to provide uniform photometry for all VIMOS
spectroscopic sources thanks to the calibrated VIMOS pre-imaging.

Follow-up priorities were assigned according to their luminosities in
order to maximize the signal-to-noise ratio (S/N herafter) and the
effective survey area within the given observation time and positional
constraints set by multi-object spectrographs. We present the sky
positions of the objects in our survey field in the left panel of
Fig~\ref{specs_pos_type}. Furthermore, the right panel of
Fig.~\ref{specs_pos_type} shows a colour-colour diagram of the
galaxies in the WFI field which we used to selected the candidates for
the VIMOS spectroscopic follow-up. 
Our aim to cover all available space on the slit masks led to the
selection of additional sources that appeared as bright ellipticals in
cases where color-selected sources could not be targeted due to typical
technical restrictions of multi-object spectrographs. Due to bad weather,
the spectroscopic follow-up shows an artificial elongation along the
southeast-northwest direction as not all scheduled VIMOS pointings in
the cluster field were performed.

\section{Optical spectroscopic data analysis and results}

\subsection{Spectroscopic data and data reduction}
\label{sec:spec_data}
The spectroscopic follow-up of potential member galaxies of S1101 were
conducted under ESO proposal ID 087.A-0096 (PI: Zhang).  We applied
for the wide-field and survey-probed capabilities of VIMOS at the UT3
(Melipal) of the ESO VLT, and were granted 22.5~hours in total for S1101
and A2597, in which only the observations of S1101 were almost completed
for $\sim$11.5~hours. The field-of-view of VIMOS is split into quadrants
of one single CCD each, \(4\times~7'\times8'\), with small gaps between
the quadrants. Using spectroscopic mode, multi-object slit masks were
manufactured for seven pointings which cover the cluster field of S1101
out to the virial radius.
All spectroscopic data for this program were taken between June and
September 2011. Typical science exposure times were 1150s and between
three and nine individual exposures per pointing were
taken. 
We used the HR-Blue grism of VIMOS in order to achieve a redshift accuracy
high enough to identify cluster substructures and obtain a reliable
$\sigma$ estimate and its deviation from the Gaussian distribution
\citep[e.g.][]{Maurogordato2008}. We measured the full width at half
maximum (FWHM) of unblended emission lines in arc-lamp exposures,
taken with the same mask and hence slit-width as the science frames.
Latter information allowed inferences on the spectral resolution,
\(R\equiv{\lambda}\cdot{\Delta\lambda}^{-1}\). Our chosen setup falls
within \(1400\lesssim R \lesssim 2070\), considering the
instrumental wavelength-range of 4200\AA~--~6200\AA~and central \(1\arcsec\) slits.
The quality of the reduced spectra allows us to not only detect emission
lines, but due to their high S/N also prominent absoption features of
early type galaxies, e.g. CaII H\&K (3934\AA~\&~3969\AA), H$\beta$
(4861\AA), H$\delta$ (4103\AA), [OIII] (4959\AA~and 5007\AA) and so on.

The data were reduced within ESO-Midas \citep{Banse1983} with a suitably tailored set of
scripts bundled together for a largely automatic pipeline reduction.
The reduction workflow includes the following tasks: de-biasing of raw
input frames, flat-fielding and cosmic filtering of scientific frames,
wavelength calibration, weighted extraction of 1-dimensional (1D)
spectra \citep[following][]{Horne1986} including generation of respective
error spectra, creation of 2-dimensional (2D) sky-subtracted frames and,
finally, co-addition of single flux-calibrated spectra.

We were able to extract a total of 492 spectra, of which 457 were of
sufficient quality for our scientific use.  We initially examined all
spectra using EZ \citep[][]{Garilli2010} to get
the initial measure of the quality and redshift of each reduced spectrum.
For spectra within a generous range around the cluster redshift (\(0.04\le z\le
0.08\)) the final redshift was computed as average of independent
gaussian fits to available spectral features in the individual spectra
to sustain high accuracy for tentative cluster members. 
The CaII H\&K (3934\AA~\&~3969\AA) absorption lines
for passive galaxies were fitted primarily.
In addition we used the G-band (4304\AA), H$\delta$
(4103\AA), H$\gamma$(4340\AA), H$\beta$ (4861\AA), and
MgII (5175\AA) absorption features, as well as [OII]
(3726\AA/3729\AA) and [OIII] (4959\AA~and 5007\AA) emission lines,
when permitted by the spectral range and feature quality.

We chose to correct our individual redshifts with respect to the
barycentre of the solar-system, in order to reduce the systematics
when averaging multiple observed sources and combining with galaxy
samples from the literature, as described in the next subsection. By
comparing acquisition images of slit masks with VIMOS pre-imaging
data, we were able to allocate individual redshifted spectra to object
coordinates from the Guide Star Catalogue-II. We take the unweighted
average of the measured redshifts for objects observed multiple times.

The accuracy of our redshift determination is affected
by the accuracy of the wavelength calibration, the precision with which
emission and absorption lines can be determined and by the scatter of
repeated redshift measurement for a given galaxy. These were added
quadratically and result in a typical redshift error of
\(\Delta z\sim0.00007\) (\(\sim 20~\rm km~s^{-1}\)), a value also found
by \citet{Maurogordato2008} using the same spectroscopic setup.

The VIMOS spectroscopic sample comprises 392 individual 
objects. According to the spectral classification, there are 220 active
galaxies (with a clear indication of emission lines), 129 passive
galaxies, 42 most likely late-type stars, and one
Lyman-\(\alpha\) emitter at redshift
\(\sim3.4\). A redshift histogram of their redshift distribution,
omitting the Lyman-\(\alpha\) emitter, is given in
Fig.~\ref{campaign_histo}.

\subsection{Incorporating public redshifts to our data}
\label{prop_catalog}
We queried the NASA/IPAC Extragalactic Database (NED) and merged the
20 redshifts in the literature
\citep[i.e. ][]{DeVaucouleurs1991,Shectman1996,Zabludoff1998,Jones2009}
from objects at the tentative cluster redshift to our sample. We
observed a scatter in redshift between galaxies from the literature
and re-observations from this study but no systematics for the
individual publications were found. Since not all sources in the
literature have published quality assessments of their spectra, we use
our redshifts instead of their NED counterparts,
literature redshifts complementary to our work (e.g. outside the field
covered by the VIMOS spectroscopy) were included.

\begin{figure}[t]
 \includegraphics[width=9cm]{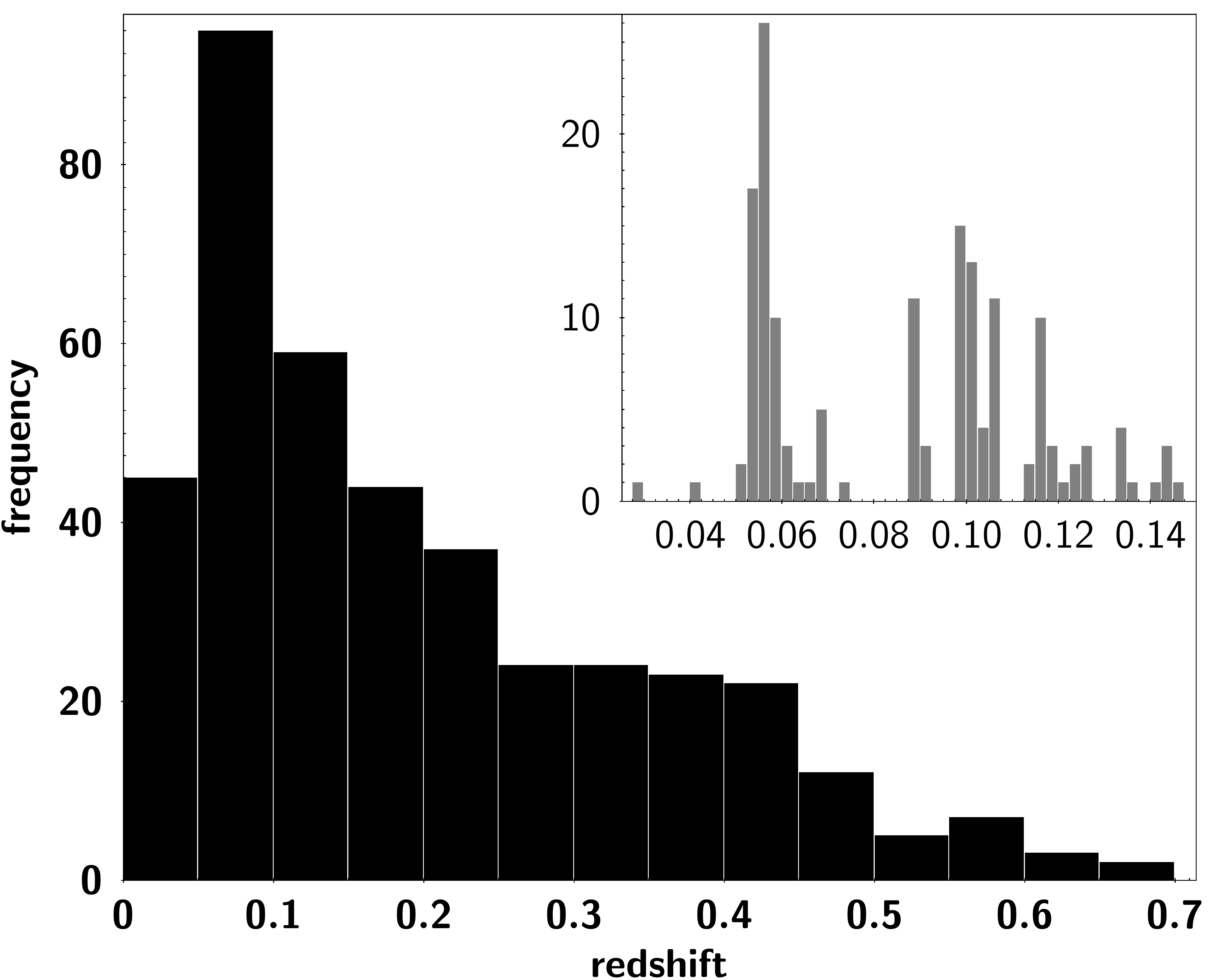}
 \caption{Redshift histogram of the VIMOS spectroscopic sample
 excluding the Lyman-\(\alpha\) emitter at \(z\sim3.4\).
 The inset in the upper right allows a closer inspection of a
 redshift window around S1101 at \(z\sim0.056\). Here, a large
 fraction of field galaxies as well as a possible background
 structure at \(z\sim0.1\) in our sample become recognizable.}
 \label{campaign_histo}
\end{figure}

Within the range of \(0.05 \le z \le 0.065\) our redshift catalogue
comprises 61 distinct galaxies (including 11 sources from the NED). We
describe the cluster member
selection in the following subsection.
Additionally, a rather large number of background galaxies in the pool
of our spectra at redshifts of around twice the cluster redshift (see
Fig.~\ref{campaign_histo}) were detected. This apparent overdensity of
galaxies at \(z\sim0.1\) is discussed further in \S~\ref{Discussion}.

\subsection{Member selection from spectroscopic data}
\label{sec:spec_member}

In order to investigate the dynamics of the cluster S1101, we need to
assign cluster membership to the galaxies in our redshift catalogue.
This is not trivial considering the fact that infalling or merging
structures exist even in relaxed systems \citep[see
e.g.][]{Zhang2012}.  We initially selected those galaxies fulfilling
\(|cz_{i} - c\overline{z}| \le 4000~\textnormal{km~s}^{-1}\) as
cluster member candidates. Here, \(cz_{i}\) is the recessional
velocity of the individual galaxy, and \(c\overline{z}\) the mean
velocity of the galaxy cluster S1101 based on its redshift in the
REFLEX Cluster Survey catalogue \citep{Boehringer2004},
\(\overline{z}=0.0564\).

\begin{figure}[t]
  \includegraphics[width=9cm]{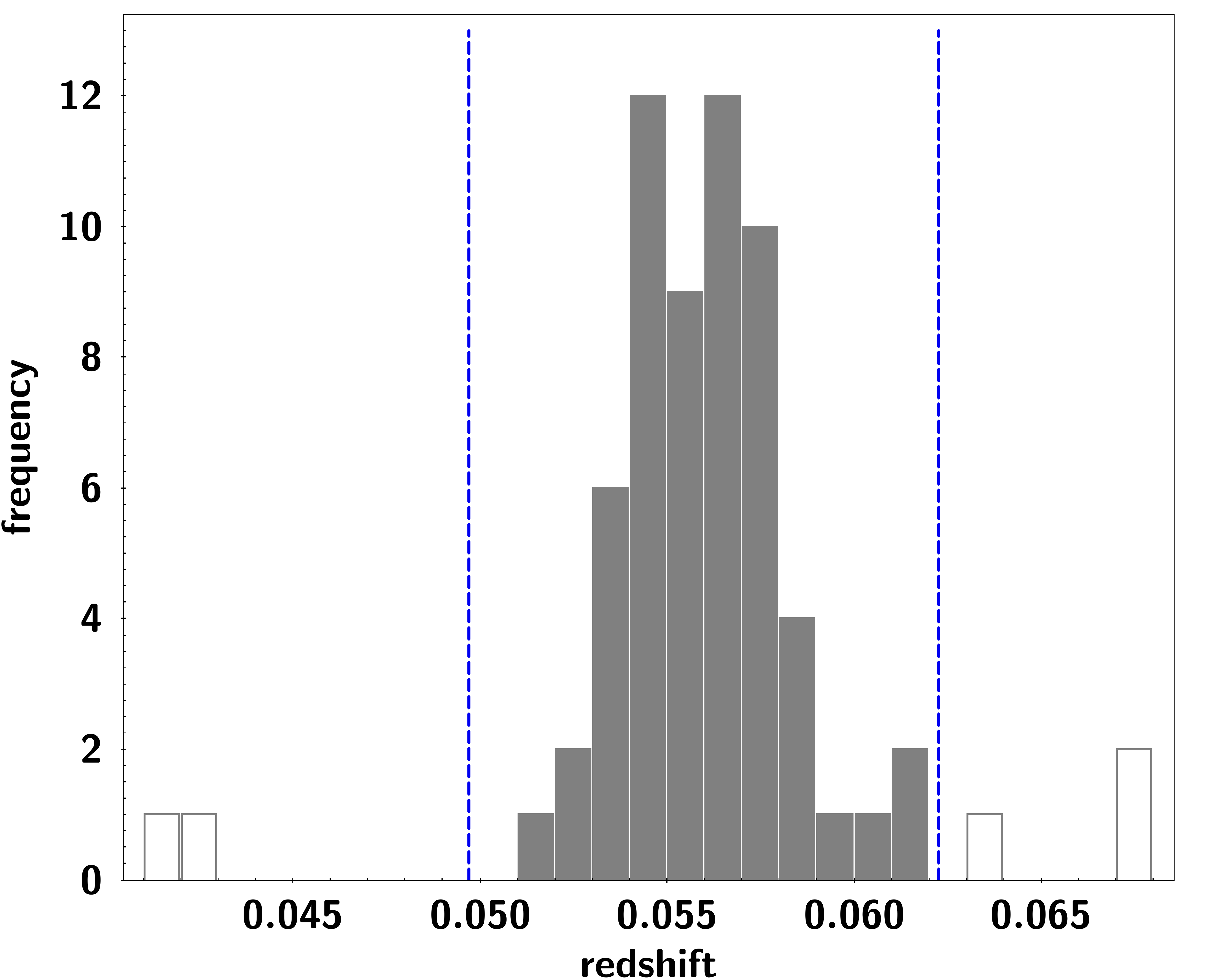} \caption{Redshift
    distribution of initially selected member galaxies in the galaxy
    cluster S1101. All visible galaxies were initially selected by
    \(|cz_{i} - c\overline{z}| \le 4000~\textnormal{km~s}^{-1}\) as
    described in \S~\ref{sec:spec_member}. Estimators of location and
    scale \citep[\(z_{cl}\) and \(\sigma\), respectively;
    see][]{Beers1990} were used to clip interlopers. The resulting
    cluster galaxy candidates are indicated by the shaded histogram,
    while the blue vertical lines refer to the \(\pm 3\sigma\)
    interval of \(\sigma_{cl}\) around \(z_{cl}\) used for the
    clipping.}
 \label{histo_cm}
\end{figure}

A revised cluster redshift (\(z_{cl}\)) and velocity dispersion
(\(\sigma_{cl}\)) was computed using the biweight estimates of
location and scale following \citet{Beers1990}.  Following
\citet{Biviano2003}, we applied the \(|z_{cl}-z_{i}| \cdot c \le 3
\sigma_{cl} \) clipping to finalize the selection of member galaxies
and repeated the computation. The resulting sample consists of 58 member
galaxies, as visualized in Fig.~\ref{histo_cm}. 

A different approach of assigning cluster membership to galaxies
is the method of ``fixed gaps``. Following the procedure by
\citet{Katgert1996} we used a gap-size of \(1000~\textnormal{km~s}^{-1}\)
and identified 59 galaxies as likely cluster members, the same 58 from
above plus one additional at \(z\sim0.0635\) (see Fig.~\ref{histo_cm}).
Since the difference in velocity dispersion (\(+5\%\)) and mean cluster
redshift (\(+0.02\%\)) are negligible, we stick to the previous procedure
with iterative clipping based on the \(\sigma_{cl}\) for the following
considerations. The equatorial coordinates and redshifts for all
spectroscopic member galaxies identified in this work are given in
Table~\ref{app_S1101} in Appendix~\ref{App_S1101}.

Fig.~\ref{specs_pos_type} shows the sky positions of all spectroscopic
member galaxies identified, together with their types being either
passive or star-forming according to their characteristic spectral
features (absorption- or emission-line dominated spectrum) once
available. 
We note that some literature sources have no publicly
available spectra for the classification.
The restriction to this simple classification was chosen
to gather information on the distribution of star forming and passively
evolving galaxies with respect to the cluster centric radii and
a possible resulting bias for the \(\sigma\) measurement.
There is an obvious overdensity of passively evolving galaxies in our
sample because we targetted the galaxy candidates selected by the
photometric colour around the red-sequence in our observations.

The final values of the cluster redshift and its velocity dispersion
following the procedure of \citet{Beers1990}
are \(z_{cl}=0.05601\pm0.00027\) and
\(\sigma_{cl}=\left(574^{+38}_{-36}\right)\textnormal{km}~\textnormal{s}^{-1}\),
respectively, are summarized in Table~\ref{tab_results} with
additional characteristic values of the cluster.
Unfortunately the fraction of active galaxies is
too low to compute an accurate \(\sigma\) to directly compare
\(\sigma_{pas}\) and \(\sigma_{act}\). The value measured
for the passive fraction, however, agrees well within the errors with the
total sample of cluster members (see Table~\ref{tab_results}).

The cluster radius $r_{\Delta}$ (e.g. $r_{500}$) was defined as the
radius within which the mass density is $\Delta$ (e.g. 500) times of
the critical density at the cluster redshift, $\rho_{\rm
  c}(z)=3H_{0}^2(8\pi G)^{-1}E^2(z)$, in which
$E^2(z)=\Omega_{M}(1+z)^3+\Omega_{\Lambda}+(1-\Omega_{M}-\Omega_{\Lambda})(1+z)^2$.
The cluster mass can be estimated by the scaling relations calibrated
by simulations from the velocity dispersion. An isothermal spherical
mass model generally does not fit well to the mass distribution of
clusters. The scaling relation based on such a model
\citep[e.g.][]{Carlberg1997a} thus provides a biased
estimate. Instead, a NFW model \citep{Navarro1997} is a better
description of the halo mass profile of clusters.
  Based on latter model \citet{Munari2013} analysed
  different simulations of group- and cluster-sized structures
  involving different physics.  Therein they estimated the velocity
  dispersion versus cluster mass relation of the following tracers of
  the cluster mass distribution: (i) dark matter particles, (ii)
  subhalos and (iii) galaxies,
\begin{equation}
 \frac{\sigma_{\rm 1D}}{\textnormal{km~s}^{-1}}=A_{\rm 1D}\left[\frac{h(z)~M_{200}}{10^{15}M_{\odot}}\right]^{\alpha}
\end{equation}
where \(\sigma_{\rm 1D}\) is the line-of-sight velocity dispersion and
\(h(z)\) the Hubble parameter at the redshift \(z\) normalized by
100~km~s\(^{-1}\)~Mpc\(^{-1}\).  \citet{Munari2013} conclude that the
best-fit values of \(A_{\rm 1D}\) and \(\alpha\) using dark matter
particles as the tracer are in general agreement with the NFW mass
profile and provide consistent results as that from
\citet{Evrard2008}.  The fits for subhalos and galaxies as tracers on
the other hand show a steeper \(\sigma\) - M relation. Since we aim
for determining the cluster mass of S1101 from the velocity dispersion
of member galaxies, we used the full range of parameters found by
tracing galaxies for the different physical models
\citep[ceased star formation and AGN feedback; compare
    Table~1 in][]{Munari2013}, \(A_{\rm 1D}=1169.75\pm11.45\) and
\(\alpha=0.3593\pm0.0068\).  The corresponding mass and radius for the
cluster S1101 are \(M_{200}=1.92^{+0.52}_{-0.42}\times
10^{14}~M_{\odot}\) and \(r_{200}=1.169^{+0.097}_{-0.093}\)~ Mpc,
given the uncertainies in \(A_{\rm 1D}\), \(\alpha\) and the error
of our velocity dispersion \(\sigma_{cl}\) (see also Table~\ref{tab_results}).
We note that the bias in using the galaxies as the tracer is low for
low-$z$ and low-mass clusters \citep[Fig.~7 in][]{Munari2013}, which
is the case of S1101.

\subsection{2D kinematic structure}
\label{sec:kinematic_structure}

The velocities of member galaxies scatter around the Hubble-flow
velocity of the underlying dark matter halo. According to the
hierarchical clustering model, clusters grow through the accretion of
galaxies and groups of galaxies falling into the cluster potential
well. Knowledge of the substructures of a galaxy cluster is however
crucial for the measurement of the line-of-sight velocity dispersion
as well as its resulting mass proxy as substructures introduce biases
in its velocity dispersion estimate \citep[e.g. ][]{Biviano2006}. The
spatial and line-of-sight velocity distributions of galaxies allow to
distinguish member galaxies taking part in an idealized gaussian
velocity distribution around the cluster mean velocity from the
infalling galaxies as well as groups of galaxies or even larger
merging events.

\citet{Pinkney1996} extensively probed various approaches of substructure
tests based solely on velocities, positions or a combination of both.
Accordingly, an advanced and robust test for substructure was described
by \citet{Dressler1988}. They provided a statistical approach to test the
2D kinematic structure of a cluster using its galaxy positions and velocities:
\begin{equation}
  \delta_{i}^{2}=\frac{N_{\rm loc}+1}{\sigma^{2}}\left[\left(\overline{v}_{i,{\rm loc}}
      -\overline{v}\right)^{2}+\left(\sigma_{i,{\rm loc}}-\sigma\right)^{2}\right].
 \label{deltaDS} 
\end{equation}
Eq.~\ref{deltaDS} makes use of \(\overline{v}_{i,{\rm loc}}\),
\(\overline{\sigma}_{i,{\rm loc}}\) and \(N_{\rm loc}\) being the mean
velocity, mean velocity dispersion and galaxy number of the local
group. Here \(\overline{v}\) and \(\sigma\) are the global mean
velocity and the velocity dispersion of the cluster. The index \(i\)
runs over all galaxies in the sample. The reduced statistics for this
test is given by
\begin{equation}
 \Delta_{DS}=\sum_{i}\delta_{i}.
 \label{DELTADS}
\end{equation}
Following \citet{Dressler1988}, the value of Eq.~\ref{DELTADS} would
exceed the number of all member galaxies (\(N_{\rm mem}\)) for systems
with substructure. We calculated \(\Delta_{DS}\) for our sample of 58
cluster member galaxies by varying group sizes, \(5\le N_{\rm loc}\le
10\). The results are listed in Table~\ref{tab_DS}. Despite
the fact that according to \citet{Pinkney1996} \(N_{\rm loc} \sim
\sqrt{N_{\rm mem}}\), for all considered values of \(N_{\rm loc}\),
the resulting \(\Delta_{DS} / N_{\rm mem}\) turned out to be well below
unity. The test therefore yielded no evidence for the presence of
substructure in the cluster S1101.

However, a method more robust that using the ``critical value'' of
unity persists in analysing the probability of \(\Delta_{\rm DS,~sim}\)
exceeding the observed \(\Delta_{\rm DS,~obs}\). According to \citet{Dressler1988}
\(\Delta_{\rm DS,~sim}\) is calculated from an artificial sample of randomly
re-assigned redshifts to the galaxy positions in the sample, the
so-called ``Monte Carlo shuffling''.
The randomisation process of the galaxy systems destroys any
true correlation between position and velocity within those simulated
clusters. Accordingly, the probabilities can be computed using
\begin{equation}
 P=\frac{f\left(\Delta_{\rm DS,~sim} > \Delta_{\rm DS,~obs}\right)}{n_{\rm sim}}~,
 \label{Pdelta}
\end{equation}
where \(\Delta_{\rm DS,~obs}\) is derived from Eq.~\ref{deltaDS} and
Eq.~\ref{DELTADS} for the observed sample, while \(\Delta_{\rm DS,~sim}\)
is computed for the sample of the reshuffled members. Only systems
with \(\Delta_{\rm DS,~sim}\) larger than the corresponding observed value
contribute to the function \(f\) in the probability \(P\) given by
Eq.~\ref{Pdelta}. For systems with clear substructures \(\Delta_{\rm
  DS,~sim} > \Delta_{\rm DS,~obs}\) is unlikely performed by randomisations
and the probability takes small values. 
Hence we consider a probability of \(P < 0.01\) (corresponding to \(<1\%\)
rejection of substructure) as a clear indication of substructure.
\begin{table}
  \caption{Results of the Dressler-Shectman test applied to the galaxy
  cluster S1101. The substructure test was performed for various group sizes
  \(N_{loc}\), and \(\Delta_{DS}\) was calculated according to Eq.~\ref{DELTADS}
  and Eq.~\ref{Pdelta} from \S~\ref{sec:kinematic_structure}.
  The listed probability \(P\) was calculated from Monte Carlo simulations
  and rejects substructure in cluster S1101 with probilities \(\geq70\%\),
  depending on the local group size.}
 \label{tab_DS}
 \centering
 \begin{tabular}{c c c }
\hline\hline\\
  \(N_{\rm loc}\) & \(\Delta_{DS}\) & \(P\) \\
  \hline\\
  5 & 30.781 & 0.882  \\
  6 & 32.986 & 0.857 \\
  7 & 40.360 & 0.768 \\
  8 & 41.710 & 0.739 \\
  9 & 42.239 & 0.720 \\
  10 & 42.999 & 0.700 \\
 \hline
 \end{tabular}
\end{table}

We calculated \(P\) using \(n_{s}=100~000\) realizations to randomly
shuffle our data for different local group sizes (5 to 10), the
results are given in Table~\ref{tab_DS}.  We find no evidence
of substructures in S1101 using either the method based on \(\Delta_{DS}\)
or the method based on \(P\). S1101 is rather relaxed compared to e.g.
the cluster R1504, for which \citet{Zhang2012} found \(P=0.06\) and
with evidence of an infall group at about \(r_{500}\).

\begin{table}
  \caption{Properties of the galaxy cluster S1101. The number in
  parenthesis for spectroscopical measures relates to the number
  of input spectra. We calculated the cluster redshift (\(z_{cl}\))
  and the velocity dispersion from all cluster members (\(\sigma_{cl}\))
  as well as from the fraction of passive galaxies alone (\(\sigma_{pas}\)),
  following \citet{Beers1990} as described in \S~\ref{sec:spec_member} and
  used the values to compute \(r_{200}\) and \(M_{200}\) following the
  NFW-fit based scaling in \citet{Munari2013}. The density normalized
  to the critical density, \(\rho_{c}(z_{cl})\), was
  estimated for all radii of Eq.~\ref{eq_caustic_mass}, and used to
  derive the radii and masses from the caustic mass profile for
  \(\Delta=500\) and 200, respectively. In the middle part X-ray
  properties are listed. \(L_{\rm bol,500}\) is the cool-core
  corrected X-ray luminosity. \(M_{gas,500}\) is the gas mass.
  \(M_{500}\) and \(r_{500}\) are the total mass and cluster radius
  derived from the gas mass distribution using the gas mass versus
  mass scaling relation. Those X-ray measurements are based on the
  combined \emph{XMM-Newton} and \emph{ROSAT} data from
  \citet{Zhang2011}. Values with diamond superscription are the X-ray
  results from \citet{Reiprich2002} using only \emph{ROSAT} pointed
  observations.}
 \label{tab_results}
 \centering
 \begin{tabular}{l r }
  \hline\hline\\
\multicolumn{2}{l}{Spectroscopic properties (58 galaxies)} \\
  \hline\\
  \(z_{cl}~(58)\) & \(0.05601\pm0.00027\) \\
  \(\sigma_{cl}~(58)\) & \(\left(574^{+38}_{-36}\right)\textnormal{km}~\textnormal{s}^{-1}\) \\
  \(\sigma_{pas}~(40)\) & \(\left(593^{+70}_{-63}\right)\textnormal{km}~\textnormal{s}^{-1}\) \\
\(r_{200}\) & \(\left(1.169^{+0.097}_{-0.093}\right)\textnormal{Mpc}\) \\
\(M_{200}\) & \(\left(1.92^{+0.52}_{-0.42}\right)\times10^{14}M_{\odot}\) \\
  \(r_{500}^{\textnormal{caustic}}\) & \(\left(0.715\pm0.005\right)~\textnormal{Mpc}\) \\
  \(M_{500}^{\textnormal{caustic}}\) & \(\left(1.10\pm0.72\right)\times10^{14}M_{\odot}\) \\
  \(r_{200}^{\textnormal{caustic}}\) & \(\left(1.033\pm0.005\right)~\textnormal{Mpc}\) \\
  \(M_{200}^{\textnormal{caustic}}\) & \(\left(1.32\pm0.93\right)\times10^{14}M_{\odot}\) \\
  \hline\hline\\
  X-ray properties \\
  \hline\\
  \(L_{\rm bol,500}\) & \(\left(1.17\pm0.10\right)\times10^{44}~\textnormal{erg~s}^{-1}\) \\
  \(r_{500}\) & \(\left(0.84\pm0.02\right)~\textnormal{Mpc}\) \\
  \(M_{500}\) & \(\left(1.87\pm0.10\right)~\times10^{14}M_{\odot}\) \\
  \(M_{gas,500}\) & \(\left(1.9\pm0.13\right)~\times10^{13}M_{\odot}\) \\
  \(r_{200}^{\diamond}\) & \(\left(1.37^{+0.26}_{-0.18}\right)~\textnormal{Mpc}\) \\
  \(M_{200}^{\diamond}\) & \(\left(2.91^{+1.99}_{-0.99}\right)~\times10^{14}M_{\odot}\) \\
  \hline\hline\\
  SZ properties \\
  \hline\\
  \(M_{500}^{SZ}\) & \(\left(4.06\pm0.92\right)\times10^{14}M_{\odot}\) \\
 \hline
 \end{tabular}
\end{table}

\subsection{Dynamical analysis using the caustic method}
\label{sec:caustic}

As mentioned in \citet{Kaiser1987}, cluster member galaxies tend to
form a ``trumpet shape'' in the project cluster centric distance
versus velocity diagram due to the proportionality between the maximum
allowed velocity, the escape velocity, relative to the cluster and the
enclosed cluster mass.  Galaxies with velocities beyond the ``trumpet
shape'' are considered to be not bound to the cluster potential well,
and are therefore not considered as cluster members. The border of the
``trumpet shape'' defines roughly the maximum velocity scatter within
the cluster, and is known as the so-called ``caustic'', an estimator
of the escape velocity for the halo matter distribution with a
spherical symmetry. A higher member statistic allows for a more
complete sampling of the caustic, which yields more robust
measurements of the underlying dark matter distribution of the cluster
\citep[e.g.][]{Diaferio1997}. The method to measure the caustic is
described in detail in e.g. \citet{Diaferio1999} and
\citet{Geller1999}. A good summary and detailed application can also
be found in \citet{Serra2011}.

The algorithm of the caustic method uses a 2D-adaptive kernel to
calculate a smoothed density distribution from the phase-space
coordinates of the input galaxies. In this context, the phase-space
spanned along the projected distance and line-of-sight velocity of the
galaxies with respect to the cluster centre and its mean velocity,
respectively. By definition, the method can be used to identify the
membership of galaxies based on the density distribution, which forms
a critical curve. In addition, the curve being an estimator of the
escaping velocity can be used to derive the mass profile of the cluster
halo. The mathematical basics of this method can be found in
\citet{Silverman1986}, the 1D-approach is outlined in \citet{Pisani1993},
wheras the multi-dimension extension is developed in \citet{Pisani1996}.

The input sample for the caustic analysis is based on all galaxies
which fulfil \(|cz_{i}-cz_{cl}| \le 4000~\textnormal{km~s}^{-1}\), where
\(z_{cl}\) is the cluster redshift calculated from the biweight estimator
of location \citep[][compare Table~\ref{tab_results}]{Beers1990}. We
note here, that in the following we use our own implementation of the
caustic code.  Nevertheless, we were fortunate to validate our results
against available routines ({\tt The Caustic App} - Ana Laura Serra \&
Antonaldo Diaferio, private communication) and remark major
differences below.

We first computed the projected radial distance of each galaxy and its
line-of-sight velocity in the cluster restframe. Contrary to the
procedures described in \citet{Serra2011} and \citet{Diaferio1999}, we
did not weight or average the galaxy positions to derive the cluster
centre but for simplicity used the BCG position (23:13:58.60, –42:43:39.0)
as the reference cluster centre. Since the BCG of the cluster S1101 is a
large and luminous central dominant galaxy and S1101 is rather relaxed, we
argue that its position is a good proxy for the global centre of
the dark matter halo --- a fact also discussed in \citet{Beers1983}.

We used the formulas from \citet{Zhang2012} to compute projected distance
and velocity in the cluster restframe: \(r=D_{a}sin\theta\) and
\(v=\left(cz-cz_{cl}\cos\theta\right)/\left(1+z_{cl}\right)\), respectively,
for all galaxies. Here, \(z_{cl}\), \(D_{a}\), \(\theta\) and \(z\)
are the cluster redshift, angular diameter distance at the cluster
redshift, angular separation from the cluster centre and galaxy redshift.
Observables and uncertainties were rescaled to match the same units (Mpc)
throughout the computation.

Now the 2D density distribution was calculated according to:
\begin{equation}
 \label{fq}
 f_{q}(\mathbf{x})=\frac{1}{N} \sum_{i=1}^{N} \frac{1}{h_{i}^{2}}K
 \left(\frac{\mathbf{x}-\mathbf{x_{i}}}{h_{i}}\right),
\end{equation}
where \(\mathbf{x}=\left(r,v\right)\)~is the vector of our input data,
whose components are weighted by the local smoothing length \(h_{i}\),
to get reasonable relative scales between radius and
velocity. \(N\) denotes the number of data points
taken into account. The index \(q\) is defined as
\(q=\sigma_{v}/\sigma_{r}\), indicates the relation between the
measurement uncertainties in velocity and radius, and will be used
later as input for Eq.~\ref{eq_h_opt}. The input coordinates have to
be rescaled in a manner that \(q\) is in the acceptable range of 10 to
50 \citep[e.g.][]{Diaferio1999}.  The kernel \(K\), is defined as:
\begin{equation}
 K(t)=\left\{
 \begin{array}{l l}  
   4\pi^{-1}\left(1-t^{2}\right)^{3}\,, & \quad t < 1 \\
   0\,, & \quad \textnormal{otherwise}
 \end{array}\right.
\end{equation}
with \(t=\frac{\mathbf{x}-\mathbf{x_{i}}}{h_{i}}\).  To derive
absolute values for the smoothing length, we start with \(h_{i} =
h_{c}h_{opt}\lambda_{i}\). Here \(\lambda_{i} =
\sqrt{\gamma~f_{1}\left(\mathbf{x_{i}}\right)^{-1}}\), with
\(\log\gamma = \sum_{i} N^{-1}\log f_{1}\left(\mathbf{x_{i}}\right)\),
where \(f_{1}\) can be derived from Eq.~\ref{fq} by fixing \(h_{i}\)
for all \(i\) to the optimal smoothing length
\begin{equation}
 \label{eq_h_opt}
 h_{opt} = \frac{6.24}{N^{1/6}}\sqrt{\frac{\sigma_{r}^{2}+\sigma_{v}^{2}}{2}}.
\end{equation}

The factor in Eq.~\ref{eq_h_opt} is different in the literature,
here we adopted the value from \cite{Serra2011}.

A minimization of
\begin{equation}
 \label{M0}
 M_{0}(h_{c}) = \frac{1}{N^{2}}\sum_{i=1}^{N}\sum_{j=1}^{N}
 \left(\frac{1}{h_{j}^{2}}K\right)^{2}~-~\frac{2}{N(N-1)}
 \sum_{i=1}^{N}\sum_{j \ne i}\frac{1}{h_{j}^{2}}K
\end{equation}
returns the smoothing parameter \(h_{c}\). The returned value works
fine, in general, with well sampled cluster populations. Sub-optimized
scaling can, however, result in global under- or over-estimation of
the contribution from the considered galaxies, leading to an either
too fine or over-smoothed density distribution \(f_{q}\) that biases
the estimation of the escape velocity and thus mass. The minimization
gives \(h_{c}=0.5682\) according to our data.

\begin{figure*}
 \centering \begin{minipage}[h]{17.5cm}
 \includegraphics[angle=0,width=17.5cm]{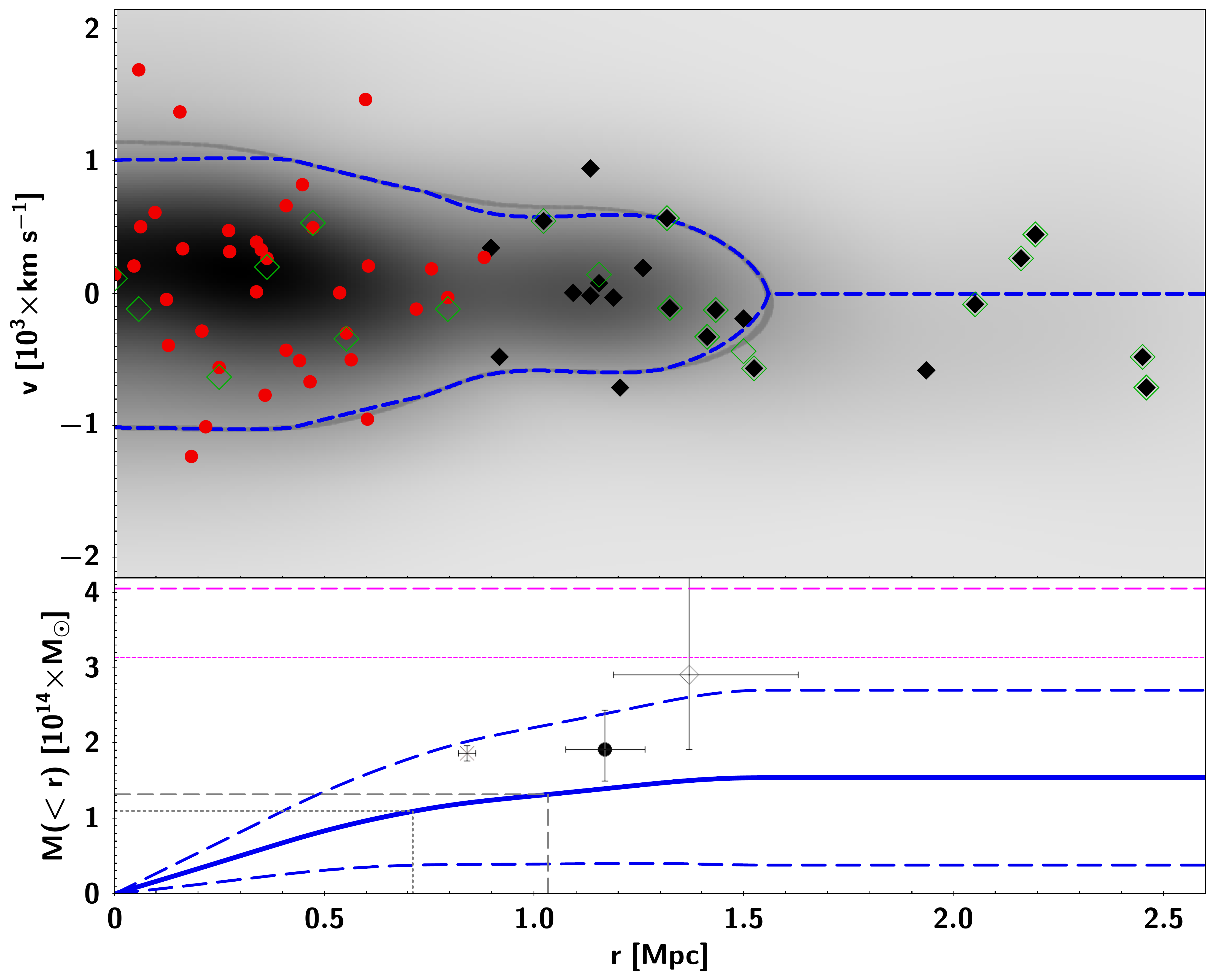}
\end{minipage} \caption{\emph{Upper panel:} Diagram of relative
  line-of-sight velocity versus projected cluster centric distance of
  the candidate member galaxies in S1101. Grey shadow shows the
  \(f_{q}(r,v)\) contours. The thick grey curve highlights the
  equal-density curve of \(f_{q}(r,v)=\kappa\). Red circles and black
  diamonds indicate galaxies within and beyond \(r=0.896\)~Mpc (used
  for constraining \(\kappa\)), respectively.  The blue dashed curve
  refers to the estimated caustic of the cluster, hence all galaxies
  with their velocities beyond are excluded from the member sample in
  this method. Green open diamonds highlight cluster members
  previously known in the literature.  \emph{Lower panel:} Mass
  profile of S1101. The mass profile from the caustic method (see
  \S~\ref{sec:caustic}) is shown as a blue thick curve, where the blue
  dashed curves show the corresponding \(1\sigma\) error estimate. The
  grey dashed and dotted lines trace the measured positions of
  \(r_{\Delta}^{caustic}\) and \(M_{\Delta}^{caustic}\) for
  \(\Delta=\{500,200\}\), respectively, according to the caustic mass
  distribution. For comparison, we overplot the dynamical mass
  \(M_{200}\) derived from the velocity dispersion according to Munari
  et al. (2013) as large black dot, the X-ray mass estimates
  \(M_{500}\) \citep{Zhang2011} and \(M_{200}^{\diamond}\)
  \citep{Reiprich2002} as black X and open diamond, respectively, as
  well as the SZ mass proxy as a dashed line with its lower error
  interval as dotted lines in magenta. Values of all mass proxies are
  also given in
  Table~\ref{tab_results}.}  \label{caustics_and_mass}
\end{figure*}

Following the procedure of \citet{Diaferio1999}, the position of the
caustic in the \(\left(r,v\right)\)-diagram is constrained by
\(f_{q}\left(r,v\right)=\kappa\). We then find the proper \(\kappa\)
by solving
\begin{equation}
 \label{eq:SkR}
 S(\kappa,R)\equiv \langle v_{esc}^{2}\rangle_{\kappa,R}-4\langle v^{2}\rangle_{R}=0~~.
\end{equation}
Eq.~\ref{eq:SkR} expresses the balancing of the escape velocity and the
velocity dispersion from galaxies within the radius \(R\), justified
by the assumption of virial equilibrium and quasi-isotropic velocity
field in the inner part of the cluster.  We encompass
\(R=0.8958~\textnormal{Mpc}\) and \(\langle v^{2} \rangle =
476.23~(\textnormal{km}~\textnormal{s}^{-1})^{2}\) from binary tree
calculations of our input data as measures for the mean radius and the
squared velocity dispersion of the galaxies associated with the
cluster branch of the binary tree. For descriptions of the binary
tree, we refer to \citet{Serra2011}.  The mean escape velocity in
Eq.~\ref{eq:SkR} remains the only \(\kappa\)-dependent quantity and can
be derived from \(\langle v_{esc}^{2}\rangle_{\kappa,R} = \int_{0}^{R}
\mathcal{A}_{\kappa}^{2}(r)\varphi(r)\textnormal{d}r~/\int_{0}^{R}
\varphi(r)\textnormal{d}r\).  We use \(R\) from the binary tree,
\(\varphi(r)=\int f_{q}(r,v)\textnormal{d}v\) and the caustic
amplitude
\(\mathcal{A}_{\kappa}(r)=\textnormal{min}\{|v_{u}|,|v_{l}|\}\).
{\(v_{u}\) and \(v_{l}\) are the upper and lower
  velocity boundary solutions of \(f_{q}\left(r,v\right)=\kappa\)
  (which could be determined to \(\kappa=0.0016461\)) for \(r\)
  running from 0 to \(R\). The cluster mass within an enclosed radius
  can be derived from the caustic amplitude as
\begin{equation}
 \label{eq_caustic_mass}
 GM(\le r)=\mathcal{F}_{\beta}\int_{0}^{r}\mathcal{A}^{2}(x)\textnormal{d}x~\textnormal{,}
\end{equation}
\ with \(G\) denoting the gravitational constant and
\(\mathcal{F}_{\beta}\) the velocity
anisotropy. Following \citet{Serra2011}, we assume
the velocity anisotropy to be independent on the radius, as a
first-order approximation, and hence chose
\(\mathcal{F}_{\beta}=0.7\).  An expression for the error of the mass
estimate was computed using \(\delta M_{i}=\sum_{j=1}^{i}
|2m_{j}\delta \mathcal{A}(r_{j}) / \mathcal{A}(r_{j})| \), where
\(\delta \mathcal{A}(r) / \mathcal{A}(r)=\kappa /
\textnormal{max}\{f(r,v)\}\) is the relative error and \(m_{j}\) the
mass of a shell at fixed radius.

The caustic mass estimates of the cluster S1101 for \(r_{500}^{\rm
  caustic}\) and \(r_{200}^{\rm caustic}\) are
\(M_{500}^{\textnormal{caustic}}=\left(1.10\pm0.72\right)\times10^{14}M_{\odot}\)
and
\(M_{200}^{\textnormal{caustic}}=\left(1.32\pm0.93\right)\times10^{14}M_{\odot}\),
respectively, and are also listed in Table~\ref{tab_results}. The
lower part of Fig.~\ref{caustics_and_mass} presents the computed mass
distribution out to a cluster-centric radius of
\(\sim1.6~\textnormal{Mpc}\). In the top panel of
Fig.~\ref{caustics_and_mass} we show the density distribution
\(f_{q}\) as grey-scale and the critical curve (\(f_{q}=\kappa\)), as
thick grey contour. The critical curve shows the caustic (blue dashed
curve) that defines member galaxies as those in it.  Galaxies with
their projected distances below or above \(R=0.8958~\textnormal{Mpc}\)
are displayed as red dots or black diamonds, respectively. We note
that the literature values included in the study (highlighted as green
diamonds) appear as a box-shape (a constant velocity dispersion
approximately along the radial distance) in the diagram, explaining
the under-estimation of the velocity dispersion in \citet{Zhang2011}
for the cluster S1101. The increased sample of cluster members in this
work helps to discover the ``trumpet shaped'' caustic expected for a
relaxed galaxy cluster. The mean values of the caustic mass,
\(M_{200}^{\textnormal{caustic}}=\left(1.32\pm0.93\right)\times10^{14}M_{\odot}\),
and the mass from the velocity dispersion based on the scaling,
\(M_{200}=\left(1.92^{+0.52}_{-0.42}\right)\times10^{14}M_{\odot}\),
(see also Table~\ref{tab_results} or lower panel of
Fig.~\ref{caustics_and_mass}) agree well within the calculated
errors. Nevertheless, the caustic mass may be underestimated
because statistical under-sampling of cluster galaxies especially at
large cluster centric radii could cause underestimation of the
caustic amplitude.

We have tested a possible bias in the mass estimate introduced from
the fact that our input data consist of only pre-selected cluster
members (see \S~\ref{sec:spec_member}), while established implementations
of the caustic code do not make assumptions on membership beforehand.
In the following, we increased the sample by including additional
galaxies from our spectroscopic galaxy sample at \(0 < z \le
1.0\). All steps computing the caustic were repeated. The resulting
density distribution is slightly different from the previous one due
to the presence of fore- and background galaxies. Nevertheless, the
caustic method still robustly identifies the same galaxies as those in
Fig.~\ref{caustics_and_mass} as the members. This confirms that the
caustic method can well distinguish cluster members from interlopers,
in which the member statistic only has impact on the precision of the
caustic mass distribution \citep[see also][]{Serra2013}. We calculated
the contrast density (\(\Delta\)) distribution, and found the cluster
radii and masses at \(\Delta=[500,200]\) vary less than 20\% between
the two samples of the input catalogues. This is reasonably caused by
the difference in the optimal smoothing and the critical value of the
density distribution.

\section{X-ray data analysis and results}

The \emph{XMM-Newton} and \emph{ROSAT} X-ray data analysis was
detailed in \citet{Zhang2011}. In the surface brightness analysis, we
directly converted the \emph{ROSAT} surface brightness profile to the
\emph{XMM-Newton} count rate using the best-fit spectral model
obtained from the \emph{XMM-Newton} data. We then combined the
\emph{XMM-Newton} surface brightness profile within the truncation
radius, where the \emph{XMM-Newton} has a S/N of $\sim 3$, with the
\emph{ROSAT} converted surface brightness profile beyond the
truncation radius for further analysis. 

The X-ray luminosity is estimated by integrating the X-ray surface
brightness. At $3\sigma$ significance, the surface brightness profile
was detected beyond \(r_{500}\) combining \emph{XMM-Newton} and
\emph{ROSAT} data. In practice, we estimated the total count rate from
the background-subtracted, flat-fielded, point-source-subtracted, and
point-spread-function corrected surface brightness profile in the
0.7--2 keV band. We converted this to X-ray luminosity using the
best-fit ``mekal'' model in XSPEC of the spectra extracted in the
aperture within the \emph{XMM-Newton} field-of-view defined in \S~3.2
in \citet{Zhang2009}. Combined observations of \emph{ROSAT} and
\emph{XMM-Newton} in Zhang et al. (2011) yield consistent total X-ray
luminosity measurements as that based on \emph{ROSAT} observations
alone in \citet{Reiprich2002} within 15\% level as shown in Fig.~A.1
in Zhang et al. (2011). This 15\% difference is a result of
introducing the corrections for point sources and substructures in
Zhang et al. (2011).

Note that the value in this work is a cool-core corrected measurement
(\(L_{\rm
  bol,500}=\left(1.17\pm0.10\right)\times10^{44}~\textnormal{erg~s}^{-1}\))
as defined in \S~2.2.2 in \citet{Zhang2011}, which is thus much lower
than the total X-ray luminosity. With the cool-core corrected
measurements we can suppress the scatter due to the cool core, which
allows us to focus on the scatter due to other facts in studying the
\(L_{\rm bol} - \sigma\) scaling relation.

As detailed in \S~2.2.1 in \citet{Zhang2011}, the cluster radius
\(r_{500}\) was determined from the X-ray gas mass profile through the
gas mass versus total mass scaling relation under the assumption of
spherical symmetry. \citet{Pratt2009} showed a tight scaling between
gas mass and total mass for a sample of 41 groups and clusters:
\begin{equation}
 E(z)^{\frac{3}{2}}\textnormal{ln}\left(\frac{M_{gas,500}}{M_{500}}\right)
 = -2.37 +0.21\textnormal{ln}\left(\frac{M_{500}}{2\times10^{14}M_{\odot}}\right)
\end{equation}
We infer the total mass profile of S1101, \(M(<r)\), from the measured
gas mass profile, \(M_{gas}(<r)\), to find the radius, \(r_{500}\),
which fulfils
\begin{equation}
 M_{500}=\frac{4\pi}{3}500\rho_{c}\left(z\right)r_{500}^{3}
\end{equation}
The resulting cluster radius is
\(r_{500}=\left(0.84\pm0.02\right)\textnormal{Mpc}\), within which the
gas mass
is \(M_{gas,500}=\left(1.90\pm0.13\right)\times10^{13}M_{\odot}\)
and the total cluster mass is
\(M_{500}=\left(1.87\pm0.10\right)\times10^{14}M_{\odot}\). Note here
that the errors of \(r_{500}\) and \(M_{500}\) are only based on the
error of \(M_{gas,500}\) which do not include the intrinsic scatter of
the scaling relation.

For comparison in Fig.~\ref{caustics_and_mass}, we also
include \(M_{200}^{\diamond}\) (see also Table~\ref{tab_results})
with an open diamond symbol, as X-ray mass estimate under the assumptions of
spherical symmetry and hydrostatic equilibrium, derived
by \citet{Reiprich2002} from \emph{ROSAT} pointed observations. 
It's not clear that the \citet{Reiprich2002} mass overestimates
the true mass since only a single beta model was fitted to the surface
brightness profile. If the profile actually steepens at large radii
then this effect may roughly compensate the isothermal assumption.
The X-ray measurements discussed above are also listed in
Table~\ref{tab_results}.  Additionally we plot
these radii and masses in the lower panel of
Fig.~\ref{caustics_and_mass} for comparison. When considering at least
10\% scatter in the gas mass versus total mass relation
\citep[e.g.][]{Okabe2010}, the mass, \(M_{500}\), from
\citet{Zhang2011} agrees rather well with the virial mass.  The mass,
\(M_{200}^{\diamond}\), from \citet{Reiprich2002} based on an
isothermal gas model slightly overestimates the mass, but still agrees
with the virial estimate \(M_{200}\) within the uncertainties
(Table~\ref{tab_results}).

\section{SPT detection of S1101}

Modern surveys with millimeter or sub-millimeter telescopes are able
to measure tiny deviations in the cosmic microwave background (CMB).
When photons from the CMB undergo inverse-Compton scattering while
passing by the intracluster medium (ICM) of galaxy clusters, an effect
known as the SZ effect (Sunyaev \& Zel'dovich 1972), a relative
temperature difference with respect to the mean CMB temperature is
expected.  This effect has a strong correlation with the mass of the
ICM and puts no redshift dependent luminosity constraints on the
observability, as for example in the X-rays. Theoretically, SZ effect
based observations are able to detect a mass limited sample of galaxy
clusters.

The South Pole Telescope \citep[SPT;][]{Carlstrom2011} is a 10m
telescope in Antarctica, and observes the southern sky in three bands
centred at 95, 150 and 220~GHz. We found S1101 in their public
catalogue of the 2500 deg\(^{2}\) data \citep[detailed in
][]{Bleem2015}. It is listed as SPT-CL J2313-4243 with a mass derived
from the SZ detection significance using the scaling relation,
\(M_{500}^{SZ}=\left(4.06\pm0.92\right)\times10^{14}M_{\odot}\).
which is much higher than the caustic as well as X-ray masses. Note
that the scaling relation between the SZ detection significance and
total mass is not particularly tight for deriving a robust mass
estimate.  Indeed, the provided \(M_{500}^{SZ}\) for S1101 is much
larger than the caustic as well as the X-ray masses, still
\citet{Bleem2015} consider masses derived for low redshift clusters
possibly biased low. We list the mass measurement in
Table~\ref{tab_results} and plotted it as a dashed magenta line in
comparison with the caustic mass profile of S1101 in
Fig.~\ref{caustics_and_mass}. 

\begin{figure}[t]
 \includegraphics[width=9.cm]{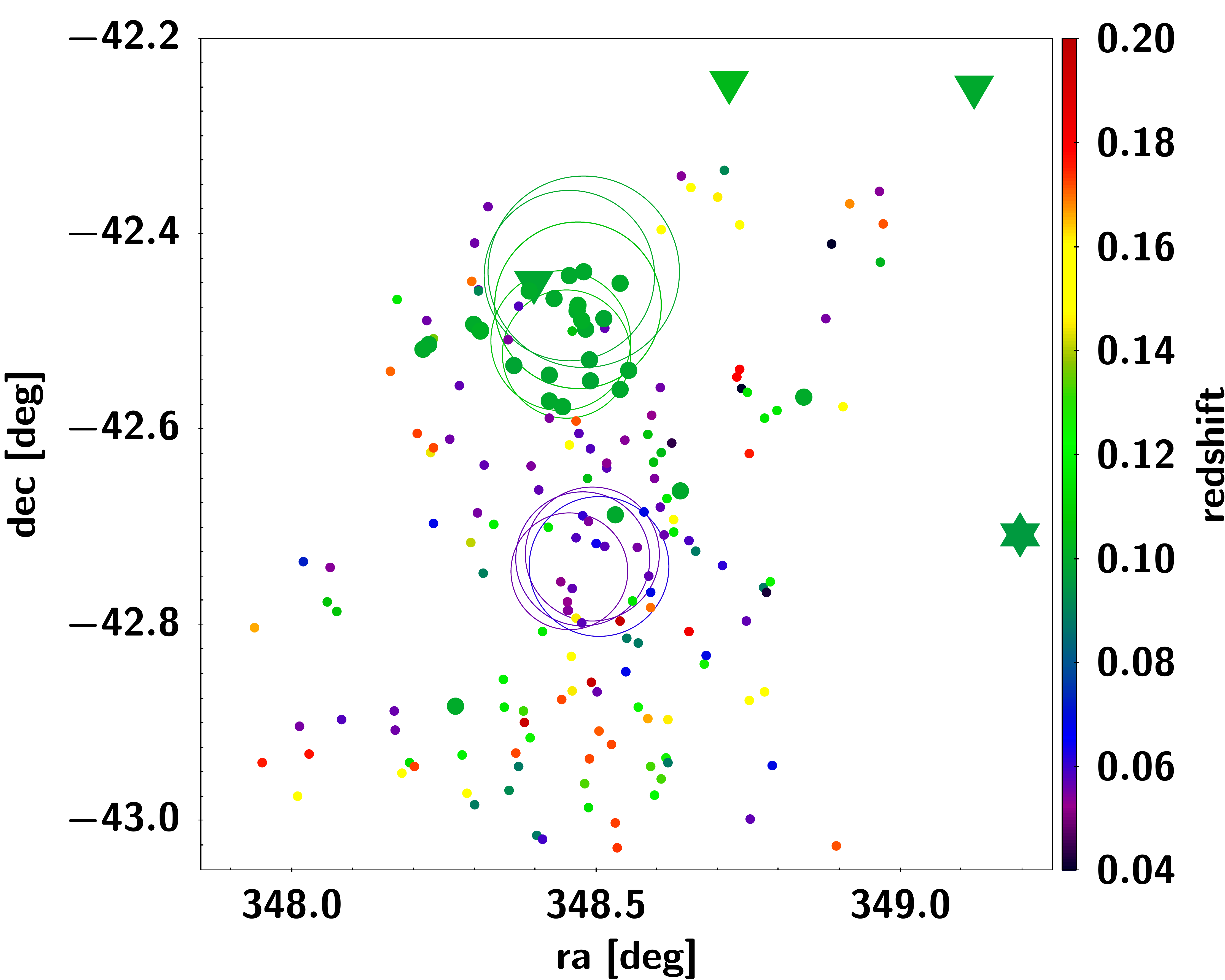}
 \caption{Dressler-Shectman plot of our spectroscopic sample {\tt S0}.
 Symbol size scales with \(exp(\delta_{i}^{2})\) above a threshold of
 \(\delta_{i} = 3\), below that value galaxies are shown as simple dots.
 Filled triangles and the filled star symbolize the central regions of
 galaxy groups and one galaxy cluster found in the vicinity of our survey
 field, as listed in the NED. The colour of all symbols corresponds to the
 redshift of the respective source.}
 \label{DS_substructure}
\end{figure}

\section{Discussion on background-structure}
\label{Discussion}
As mentioned in \S~\ref{sec:spec_data}, our spectroscopic dataset displays
an overdensity of background galaxies at approximately twice the
cluster redshift (see Fig.~\ref{campaign_histo}). We thus examined the
spectroscopic redshift catalogue comprising 191 galaxies at \(0 < z <
0.2\) (from now on {\tt S0}), which encloses cluster members and
possible background structures. In order to detect substructures
within this broad redshift range, we applied the DS-test, as described
in \S~\ref{sec:kinematic_structure}, with \(N_{loc}=10\) to this
catalogue.

In Fig.~\ref{DS_substructure}, we show the sky-positions of these
galaxies, where their symbol-sizes scale with \(exp(\delta_{i}^{2})\).
There are two clumps with high probabilities of being substructures.
As highlighted by the colour-coded redshift information, the blue
circles represent the dense core of galaxy cluster S1101, while yellow
and orange circles indicate a potential background-structure at
\(0.09 < z < 0.11\).

We made an attempt to assign membership of the galaxies to the tentative
background-structure using the following procedure:
\newline1) We selected the galaxies with the highest
\(\delta_{i}\) at larger redshift than the cluster S1101
(compare colour-coding in Fig.~\ref{DS_substructure}) and calculated
the mean redshift, \(\overline{z}\), of those 7 galaxies.
\newline2) With a gap clipping of \(|c\overline{z}-cz_{S0}| \le 4000~\textnormal{km~s}^{-1}\)
around \(\overline{z}\), a first outlier-rejected sample
called {\tt S1} was derived.
\newline3) Application of the biweight estimators combined with 
the \(|cz - c\overline{z_{S1}}| \le 3 \sigma_{S1}\) clipping in
iterative manner resulted in the final sample of 27 galaxies
for this background structure, the {\tt S2} sample.
\newline4) Final redshift and velocity dispersion of the {\tt S2}
sample  were computed using the biweight estimators of location and
scale.  Note that these steps follow the same method we applied when
selecting member galaxies for the cluster S1101 (compare \S~\ref{sec:spec_member}).

\begin{figure}[t]
 \includegraphics[width=9cm]{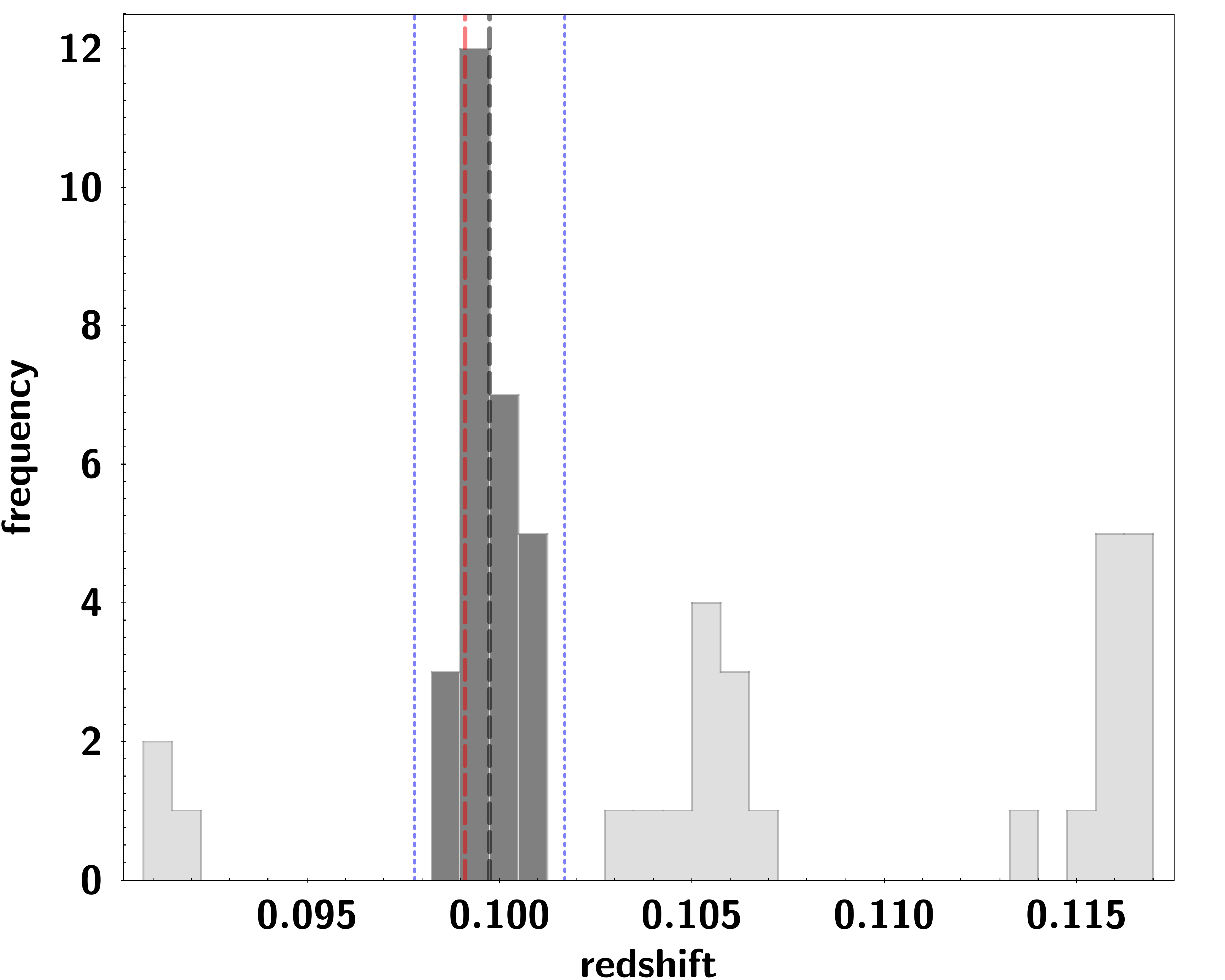}
 \caption{Redshift histogram of the sample S1 around
          \(z\sim 0.1\).
          The dotted blue vertical lines refer to the borders of the final clipping,
          while the actual members of the sample S2 are shown as dark
          shaded region. The redshift of galaxy group LCLG -42 230 is indicated
          by a black and red vertical line, based on our sample
          {\tt S2} and previous work of \citet{Tucker2000}, respectively.
          The definition of S1 and S2 are given in \S~\ref{Discussion},
          the corresponding data are listed in Appendix \ref{App_substructure}.}
 \label{LSS_histo}
\end{figure}

We note that by using the fixed gapper method with the typical gap of
\(1000~\textnormal{km~s}^{-1}\), the whole sample {\tt S1} is regarded
as a group or cluster. Refering to Fig.~\ref{LSS_histo}, additional to
the objects within the blue dotted lines (the final clipping for
sample {\tt S2}) the overdensity of 7 galaxies centred at \(z\sim0.105\)
would contribute to the member population as well. The fixed gap may,
however, be too large for such low mass systems. The additional
overdensity may indicate some structure that may be connected with but
not directly belong to {\tt S2}. The possibility of both systems
interacting gravitationally is still possible but not part of this
discussion.

For the sample {\tt S2} we derive \(z_{S2}=0.09974 \pm 0.00017\) and
\(\sigma_{S2}=\left(195^{+49}_{-39}\right)\textnormal{km~s}^{-1}\).
Following \citet{Munari2013}, we obtained a mass of
\(M_{200}=\left(0.93^{+1.00}_{-0.49}\right)\times10^{13}M_{\odot}\),
which indicates that {\tt S2} is a galaxy group, despite of its high
number of member galaxies.

A query in the NED for group or cluster counter-parts resulted in one
cluster and several groups of galaxies near the cluster S1101. The
cluster is visualized as a yellow star, while the groups are shown as
filled triangles in Fig.~\ref{DS_substructure}.  The galaxy group that
coincides best with our findings is LCLG -42 230 from the ``Las
Campanas Loose Group'' catalogue and has a published spectroscopic
redshift of 0.0991 \citep[][]{Tucker2000}.  In the Las Campanas
Redshift Survey, latter redshift was calculated from 6 galaxies only,
compared to 27 galaxies for this work. We measured a sky-distance of
\(\sim3.69\arcmin\) between the {\tt S2} mean coordinates and the
position in the LCLG catalogue, corresponding to a distance of
\(\sim406~\textnormal{kpc}\) at the redshift of the group.
This distance is comparable to the virial radius of {\tt S2}.
The 27 galaxies serendipitously found in our data may well belong
to LCLG -42 230.

The scaling relation in Fig.~\ref{lbsigma} yields an X-ray luminosity
of \(\sim4\times10^{41}\textnormal{erg~s}^{-1}\) for this group from
its measured \(\sigma\). This corresponds to a flux more than three
magnitudes lower than that of S1101. It can therefore not cause any
significant bias in the X-ray measurement of S1101.

\section{Conclusions}

We performed multi-object spectroscopic follow-up observations of the
galaxies in the S1101 field in order to increase the member statistic
for its dynamical study. The combined sample of our survey and the
redshifts from the NED yields 58 or 42 cluster member galaxies, using
the member identification method of \citet{Beers1990} or the caustic
method, respectively.

The fraction of 47 confirmed (from our VIMOS spectra
only) versus 392 candidate cluster members, corresponding to
(\(\sim 12\%\)), might appear relatively low. Considering
galaxies with magnitudes \(R<18~\rm mag\) the fraction between
confirmed and observed sources still is 24 versus 69 (\(\sim 35\%\)).
Observational restrictions demanded to also include fainter targets but
increased the influence of photometric uncertainties.
Despite unfavourable conditions during pre-imaging a sound sample of member
galaxies could be established by making advantage of the survey capabilities
offered by VIMOS.
Large MOS capabilities will however not necsessarily be available for
future large-scale automized surveys underlyning the importance of a
suitable pre-selection.

The 2D dynamical substructure tests indicate no ongoing merging
activities in S1101, also supported by the large fraction of passive
galaxies residing in the cluster centre and the absence of severe
inhomogeneities in the available X-ray data.

The increased number of member galaxies with spectroscopic redshifts
and the relaxed cluster structure indicated in our 2D test ensure
robust constraints on the velocity dispersion and virial mass
estimates. 
The recovery of the ``trumpet shape'' in the \(v-r\)
plot (Fig.~\ref{caustics_and_mass}) indicates an improved completeness
of the confirmed cluster population and hence a reduced bias in
\(\sigma\) with respect to the previous study of \citet{Zhang2011}.
Using the cluster velocity dispersion, \(\sigma\), we
updated the position of S1101 in the \(L_{\rm bol} - \sigma\) diagram
(Fig.~\ref{lbsigma}) with respect to the work of \citet{Zhang2011}.
We confirmed the hint in \citet{Zhang2011} based on the simulations
sample that the severe deviation of S1101 from the indicated scaling
relation therein is mostly due to the velocity dispersion measurement
with low member statistic. This result adds confidence for the
interpretation as a bias due to a small number of member
galaxies. Given the ongoing efforts for large scale cluster mass
calibration with \(\sim\)20~members/cluster, e.g. SPIDERS
(SPectroscopic IDentification of ERosita Sources) and 4MOST (4-metre
Multi-Object Spectroscopic Telescope) for the \emph{eROSITA} clusters
\citep[e.g. ][]{Merloni2012}, it is useful to determine a correction
factor based on our whole sample in \citet{Zhang2011}. A spectroscopic
follow-up of the remaining outlier clusters in our sample in the near
future would help to realize this goal.

\begin{figure}
\vskip-0.9cm
\hskip-1.75cm
 \includegraphics[width=14cm]{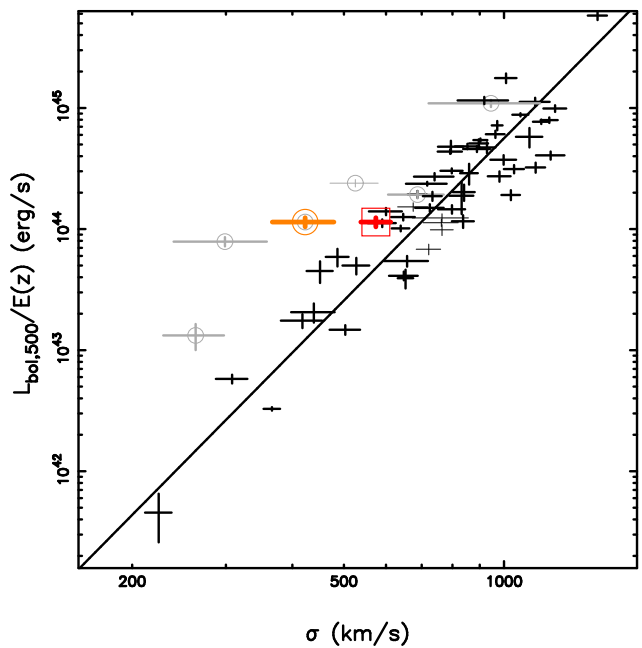}
 \vskip-1cm
 \caption{$L^{\rm cocc}_{\rm bol}-\sigma$ diagram of
     the HIFLUGCS clusters, in which the best fit is for those 56
     clusters with more than 45 spectroscopic members per cluster. The
     orange circle and red box show the original \citep{Zhang2011} and
     the revised values from this work, respectively, for
     S1101. Clusters with less than 45 spectroscopic members are shown
     as grey circles.}
 \label{lbsigma}
\end{figure}

The mass estimate from the velocity dispersion,
\(M_{200}=\left(1.92^{+0.52}_{-0.42}\right)\times10^{14}M_{\odot}\),
agrees with the available X-ray mass proxies
\citep{Zhang2011,Reiprich2002} within the uncertainties. The SZ mass
proxy \citep{Bleem2015} is much higher than the remaining mass estimates,
namely the dynamical, X-ray and caustic based masses (compare
Table~\ref{tab_results} and Fig.~\ref{caustics_and_mass}). This
might be due to the large scatter in the scaling relation between the
SZ detection significance and total mass they used. The caustic method
makes no assumption on the dynamical state of the cluster, and traces
the enclosed cluster mass as a function of radius, solely by taking into
account positions and redshifts of member galaxies. The caustic mass
profile might slightly underestimate the cluster mass given a rather
small sample of cluster galaxies distributed up to large projected
distance of 2.5 Mpc. We therefore note that the caustic-method is in
principle able to trace the clusters mass out to large radii (\(>2r_{200}\)),
but relies on high member statistic in order to reach high precision
\citep[200 or more; see][]{Serra2011}. The member statistic of S1101
in this study is still too low to deliver a high-precision application
of the caustic method.

In addition to the population of cluster galaxies, the redshift
distribution revealed an apparent overdensity at \(z\sim0.1\).  The 2D
dynamical substructure tests using galaxies at \(0 < z < 0.2\) and the
further analysis of those galaxies, suggested a group of 27 galaxies
with \(z_{S2}=0.09974 \pm 0.00017\) (see \S~\ref{Discussion}). We
measured a velocity dispersion of
\(\sigma_{S2}=\left(195^{+49}_{-39}\right)\textnormal{km~s}^{-1}\) for
this group, which yields
\(M_{200}=\left(0.93^{+1.00}_{-0.49}\right)\times10^{13}M_{\odot}\)
following \citet{Munari2013}. The resulting X-ray luminosity according
to our scaling relation is less than 1\% of that of S1101. No
significant bias due to this group should be expected in the X-ray
luminosity estimate of S1101.

\begin{acknowledgements}
  We thank an anonymous referee for constructive criticism that helped
  to improve the readability of the paper.
  We thank A.~Biviano, H.~B\"ohringer, and
  J.~Dietrich for their numerous suggestions for helping this
  work. Furthermore, we thank A.~L.~Serra and A.~Diaferio for the
  communication on the caustic method and kindness to cross-check our
  results with their code.
  The Guide Star Catalogue–II is a joint project of the Space Telescope
  Science Institute and the Osservatorio Astronomico di Torino.  Space
  Telescope Science Institute is operated by the Association of
  Universities for Research in Astronomy, for the National Aeronautics
  and Space Administration under contract NAS5-26555. The
  participation of the Osservatorio Astronomico di Torino is supported
  by the Italian Council for Research in Astronomy. Additional support
  is provided by European Southern Observatory, Space Telescope
  European Coordinating Facility, the International GEMINI project and
  the European Space Agency Astrophysics Division. We have made use of
  VLT/VIMOS observations taken with the ESO Telescope at the Paranal
  Observatory under programme 087.A-0096 and WFI observations
  partially supported by the Deutsche Forschungsgemeinschaft (DFG)
  through Transregional Collaborative Research Centre TRR 33. The
  \emph{XMM-Newton} project is an ESA Science Mission with instruments
  and contributions directly funded by ESA Member States and the USA
  (NASA). The \emph{XMM-Newton} project is supported by the
  Bundesministerium f\"ur Wirtschaft und Technologie/Deutsches Zentrum
  f\"ur Luft- und Raumfahrt (BMWI/DLR, FKZ 50 OX 0001) and the
  Max-Planck Society.  This research has made use of the NASA/IPAC
  Extragalactic Database (NED) which is operated by the Jet Propulsion
  Laboratory, California Institute of Technology, under contract with
  the National Aeronautics and Space Administration.
  We used the version 13SEPpl1.4 of ESO-MIDAS for data reduction and
  STILTS\footnote{http://www.starlink.ac.uk/stilts/} for table processing
  and plotting of figures.
  Y.Y.Z. acknowledges support by the German BMWi through the
  Verbundforschung under grant 50~OR~1506.
  T.H.R. acknowledges support from the DFG through the Heisenberg
  research grant RE 1462/5.
\end{acknowledgements}
\nocite{Taylor2006}
\nocite{Taylor2009}

\bibliographystyle{aa}

\bibliography{29043_ref.bib}

\newpage
\begin{appendix}

\section{Spectroscopic targets of LCLG -42 230}
\label{App_substructure}
This section lists sky coordinates and redshifts for the sample {\tt{S1}}, contributing to the background
structure at redshift \(z\sim0.1\) in the field of view of
S1101. The subsample, with considered affiliation
to LCLG -42 230, is highlighted as {\tt{S2}} in column 4. The sample properties are explained in detail in
\S~\ref{Discussion}, coordinates correspond to GSC2.3 sources and the listed redshifts are based on fits to
passive and active galaxy templates by {\tt{EZ}}.
\begin{table}[h]
\label{app_background}
 \caption{Sample {\tt{S1}} of the background group at \(z\sim0.1\). Note that the subsample {\tt{S2}} is
          included in {\tt{S1}} but is highlighted for simplicity as sub-group only (see column 4).
          While {\tt{S1}} corresponds to a simple velocity-clipped sample, {\tt{S2}} is based on biweight
          estimates and iterative sigma-clipping (see \S~\ref{Discussion} for details). Galaxies of latter
          sample are considered as group members of LCLG -42 230. Position in RA and DEC (J2000) is based on
          GSC2.3 astrometry. The spectroscopic redshift is given in column 3.}
\centering
\begin{tabular}{c c c c }
\hline\hline
ra & dec & z & sample \\
\hline
  23:12:13.8 & -42:46:36 & 0.1067 & {\tt{S1}} \\
  23:12:17.8 & -42:47:10 & 0.1065 & {\tt{S1}} \\
  23:12:41.7 & -42:28:04 & 0.1166 & {\tt{S1}} \\
  23:12:46.5 & -42:56:27 & 0.1134 & {\tt{S1}} \\
  23:12:51.7 & -42:31:05 & 0.1000 & {\tt{S2}} \\
  23:12:54.0 & -42:30:50 & 0.0997 & {\tt{S2}} \\
  23:13:04.7 & -42:52:59 & 0.1002 & {\tt{S2}} \\
  23:13:12.0 & -42:29:34 & 0.1002 & {\tt{S2}} \\
  23:13:13.6 & -42:27:31 & 0.0914 & {\tt{S1}} \\
  23:13:14.3 & -42:29:59 & 0.1012 & {\tt{S2}} \\
  23:13:14.5 & -42:29:57 & 0.1003 & {\tt{S2}} \\
  23:13:24.0 & -42:53:05 & 0.1167 & {\tt{S1}} \\
  23:13:25.9 & -42:58:08 & 0.0922 & {\tt{S1}} \\
  23:13:27.5 & -42:32:05 & 0.0983 & {\tt{S2}} \\
  23:13:33.6 & -42:27:33 & 0.1001 & {\tt{S2}} \\
  23:13:39.1 & -42:48:23 & 0.1164 & {\tt{S1}} \\
  23:13:41.2 & -42:42:03 & 0.1160 & {\tt{S1}} \\
  23:13:41.5 & -42:32:42 & 0.0992 & {\tt{S2}} \\
  23:13:41.8 & -42:34:15 & 0.0993 & {\tt{S2}} \\
  23:13:43.4 & -42:28:00 & 0.0994 & {\tt{S2}} \\
  23:13:46.2 & -42:30:35 & 0.1056 & {\tt{S1}} \\
  23:13:46.9 & -42:34:38 & 0.0995 & {\tt{S2}} \\
  23:13:48.4 & -42:31:24 & 0.1065 & {\tt{S1}} \\
  23:13:49.5 & -42:26:34 & 0.0993 & {\tt{S2}} \\
  23:13:50.8 & -42:30:00 & 0.1055 & {\tt{S1}} \\
  23:13:52.7 & -42:28:47 & 0.1003 & {\tt{S2}} \\
  23:13:52.8 & -42:28:23 & 0.1006 & {\tt{S2}} \\
  23:13:52.8 & -42:28:25 & 0.1062 & {\tt{S1}} \\
  23:13:54.4 & -42:29:21 & 0.1009 & {\tt{S2}} \\
  23:13:55.4 & -42:26:19 & 0.0996 & {\tt{S2}} \\
  23:13:55.8 & -42:29:51 & 0.0993 & {\tt{S2}} \\
  23:13:56.8 & -42:39:02 & 0.1044 & {\tt{S1}} \\
  23:13:57.2 & -42:59:11 & 0.1149 & {\tt{S1}} \\
  23:13:57.6 & -42:31:44 & 0.0986 & {\tt{S2}} \\
  23:13:57.9 & -42:33:02 & 0.0985 & {\tt{S2}} \\
  23:14:03.3 & -42:29:13 & 0.0992 & {\tt{S2}} \\
  23:14:07.7 & -42:41:17 & 0.0998 & {\tt{S2}} \\
  23:14:09.6 & -42:27:04 & 0.1005 & {\tt{S2}} \\
  23:14:09.7 & -42:33:33 & 0.0996 & {\tt{S2}} \\
  23:14:13.1 & -42:32:23 & 0.0995 & {\tt{S2}} \\
  23:14:14.3 & -42:46:33 & 0.1156 & {\tt{S1}} \\
\hline\end{tabular}
\end{table}
\addtocounter{table}{-1}
\begin{table}
\caption{continued.}
\centering
\begin{tabular}{c c c c }
\hline\hline
ra & dec & z & sample \\
\hline
23:14:20.5 & -42:36:21 & 0.1057 & {\tt{S1}} \\
  23:14:22.7 & -42:38:03 & 0.1051 & {\tt{S1}} \\
  23:14:25.7 & -42:37:28 & 0.1039 & {\tt{S1}} \\
  23:14:28.1 & -42:40:16 & 0.1163 & {\tt{S1}} \\
  23:14:30.9 & -42:42:17 & 0.1166 & {\tt{S1}} \\
  23:14:33.3 & -42:39:47 & 0.0994 & {\tt{S2}} \\
  23:14:50.7 & -42:20:06 & 0.0914 & {\tt{S1}} \\
  23:14:59.8 & -42:33:44 & 0.1156 & {\tt{S1}} \\
  23:15:06.7 & -42:35:18 & 0.1157 & {\tt{S1}} \\
  23:15:11.5 & -42:34:52 & 0.1158 & {\tt{S1}} \\
  23:15:22.2 & -42:34:03 & 0.1005 & {\tt{S1}} \\
  23:15:52.2 & -42:25:47 & 0.1029 & {\tt{S2}} \\
\hline\end{tabular}
  \end{table}

\section{Spectroscopic targets of S1101}
\label{App_S1101}
Using the iterative sigma-clipping method, as described in
\S~\ref{sec:spec_member}, our sample of cluster member galaxies for the
cluster S1101 comprises 58 galaxies. We list this sample, including
sky coordinates, redshift and a simple spectral classification in
Table~\ref{app_S1101}. The coordinates given in the table correspond
to GSC2.3 sources. As a reference we list spectral redshifts from
various publications (results from the NED query). Whenever data
available in our sample, we prefer to use our own for the final
redshift measurements and spectral classification (see
\S~\ref{sec:spec_data}). The visibility of the spectral emission lines is
decisive for either passive or active classification, corresponding to
{\tt{0}} and {\tt{1}} in column 4 of Table~\ref{app_S1101}. Those
objects that have no spectra available or have spectral coverage and
data quality hampered the decision process, were assigned {\tt{99}}.

\begin{table*}
\label{app_S1101}
\caption{Sample of spectroscopically confirmed 
cluster members of S1101. Columns 1--2 show object positions based on
GSC2.3 astrometry. The spectroscopic redshift is given in column
3. Column 4 refers to the spectral type: {\tt{0}} for passive spectra
without emission lines, {\tt{1}} for spectra with undoubtful [OII] or
[OIII] lines, and {\tt{99}} where poor data quality or the absence of
spectra hampered classification.
We assess a gobal uncertainty of our redshift measurements of
\(\Delta z\sim0.00007\) (or \(v\sim22~\rm km~s^{-1}\),
see \S~\ref{sec:spec_data}).
Note that 376 out of the 403 individual sources with
spectroscopic redshifts are within the field of our VIMOS
pre-imaging. Column~5 gives the $R$-band magnitude in
the Vega system for those sources from the VIMOS pre-imaging.
Column 6 lists the references of the NED
provided redshifts.}
\centering\vskip-0.25cm
\begin{tabular}{c c c c c c}
\hline\hline
ra & dec & z & type & R & reference \\
\hline
  23:11:40.7 & -43:06:31 & 0.05694 & {\tt{0}}  & --- & \citet{Jones2009} \\
  23:11:47.7 & -43:05:21 & 0.05571 & {\tt{1}}  & --- & \citet{Jones2009} \\
  23:12:03.2 & -42:54:13 & 0.05532 & {\tt{1}}  & 15.860 & --- \\
  23:12:15.2 & -42:44:28 & 0.05349 & {\tt{0}}  & 17.611 & --- \\
  23:12:19.9 & -42:53:50 & 0.05800 & {\tt{0}}  & 15.518 & \citet{Jones2009} \\
  23:12:40.5 & -42:53:16 & 0.05602 & {\tt{0}}  & 16.208 & --- \\
  23:12:41.0 & -42:54:28 & 0.05595 & {\tt{1}}  & 18.487 & --- \\
  23:12:53.1 & -42:29:19 & 0.05589 & {\tt{1}}  & 18.837 & --- \\
  23:13:02.2 & -42:36:36 & 0.05589 & {\tt{0}}  & 15.272 & --- \\
  23:13:06.0 & -42:33:21 & 0.05721 & {\tt{0}}  & 18.481 & --- \\
  23:13:12.2 & -42:24:36 & 0.05562 & {\tt{99}} & --- & \citet{Shectman1996} \\
  23:13:13.3 & -42:41:07 & 0.05495 & {\tt{0}}  & 15.013 & --- \\
  23:13:13.5 & -42:27:27 & 0.05626 & {\tt{0}}  & 16.360 & --- \\
  23:13:16.0 & -42:38:11 & 0.05673 & {\tt{0}}  & 16.195 & --- \\
  23:13:17.4 & -42:22:22 & 0.05556 & {\tt{99}} & --- & \citet{Shectman1996} \\
  23:13:25.5 & -42:30:32 & 0.05432 & {\tt{1}}  & 16.204 & --- \\
  23:13:29.7 & -42:28:27 & 0.05792 & {\tt{0}}  & 15.819 & \citet{Jones2009} \\
  23:13:34.6 & -42:38:17 & 0.05421 & {\tt{0}}  & 18.938 & --- \\
  23:13:37.5 & -42:39:42 & 0.05716 & {\tt{0}}  & 20.064 & --- \\
  23:13:38.9 & -43:01:09 & 0.05934 & {\tt{0}}  & 17.177 & --- \\
  23:13:41.4 & -42:35:20 & 0.05424 & {\tt{0}}  & 19.584 & --- \\
  23:13:46.0 & -42:45:22 & 0.05165 & {\tt{0}}  & 17.429 & --- \\
  23:13:48.8 & -42:46:34 & 0.05246 & {\tt{0}}  & 15.274 & --- \\
  23:13:48.9 & -42:47:09 & 0.05404 & {\tt{0}}  & 15.822 & --- \\
  23:13:49.6 & -42:44:44 & 0.05583 & {\tt{0}}  & 17.653 & --- \\
  23:13:50.9 & -42:45:46 & 0.05719 & {\tt{0}}  & 16.504 & --- \\
  23:13:52.3 & -42:42:39 & 0.05817 & {\tt{0}}  & 19.396 & --- \\
  23:13:53.1 & -42:36:15 & 0.05775 & {\tt{0}}  & 15.123 & --- \\
  23:13:54.6 & -42:47:53 & 0.05768 & {\tt{1}}  & 19.654 & --- \\
  23:13:54.7 & -42:41:18 & 0.06083 & {\tt{0}}  & 17.371 & --- \\
  23:13:54.9 & -42:43:57 & 0.05672 & {\tt{0}}  & 18.315 & --- \\
  23:13:57.2 & -42:41:39 & 0.05463 & {\tt{0}}  & 15.999 & --- \\
  23:13:58.0 & -42:37:11 & 0.05833 & {\tt{0}}  & 19.949 & --- \\
  23:13:58.6 & -42:43:39 & 0.05651 & {\tt{0}}  & 14.537 & --- \\
  23:14:00.3 & -42:52:07 & 0.05602 & {\tt{0}}  & 18.986 & --- \\
  23:14:01.3 & -42:44:26 & 0.06196 & {\tt{0}}  & 16.000 & --- \\
  23:14:03.4 & -42:29:47 & 0.05696 & {\tt{0}}  & 17.321 & --- \\
  23:14:03.5 & -42:43:11 & 0.05778 & {\tt{0}}  & 17.559 & --- \\
  23:14:04.3 & -42:38:05 & 0.05330 & {\tt{1}}  & 16.823 & --- \\
  23:14:04.3 & -42:38:24 & 0.05737 & {\tt{0}}  & 17.827 & --- \\
  23:14:11.3 & -42:36:40 & 0.05365 & {\tt{1}}  & 20.911 & --- \\
  23:14:16.5 & -42:43:16 & 0.05499 & {\tt{0}}  & 17.490 & --- \\
  23:14:21.0 & -42:45:01 & 0.05713 & {\tt{0}}  & 17.205 & --- \\
  23:14:22.1 & -42:35:10 & 0.05266 & {\tt{0}}  & --- & --- \\
  23:14:23.0 & -42:39:01 & 0.05449 & {\tt{0}}  & 17.238 & --- \\
  23:14:25.5 & -42:40:47 & 0.05693 & {\tt{0}}  & 15.714 & --- \\
  23:14:25.6 & -42:33:27 & 0.05558 & {\tt{1}}  & 17.465 & --- \\
  23:14:26.9 & -42:42:28 & 0.05604 & {\tt{0}}  & 16.157 & --- \\
  23:14:33.6 & -42:20:30 & 0.05402 & {\tt{0}}  & 17.025 & \citet{Shectman1996} \\
  23:14:36.8 & -42:42:51 & 0.05890 & {\tt{0}}  & 17.785 & --- \\
  23:14:49.9 & -42:44:24 & 0.06118 & {\tt{1}}  & 19.876 & --- \\
  23:14:59.5 & -42:47:46 & 0.05666 & {\tt{0}}  & 19.153 & --- \\
  23:15:00.9 & -42:59:54 & 0.05667 & {\tt{1}}  & 17.837 & --- \\
  23:15:14.7 & -42:07:40 & 0.05430 & {\tt{99}} & --- & \citet{Shectman1996} \\
  23:15:30.9 & -42:29:14 & 0.05486 & {\tt{99}}  & 16.274 & \citet{Zabludoff1998} \\
  23:15:52.0 & -42:21:24 & 0.05396 & {\tt{1}}  & 16.590 & --- \\
  23:16:31.5 & -42:23:33 & 0.05757 & {\tt{99}} & --- & \citet{Shectman1996} \\
  23:17:02.3 & -42:24:41 & 0.05349 & {\tt{0}}  & --- & \citet{Shectman1996} \\
   \hline\end{tabular}
  \end{table*}
\clearpage

\section{Properties of the combined spectroscopic sample}

We summarize the properties of all the VIMOS sources in our study
below so that the community can access them either for cross-check
or other purposes, e.g. identification of distance objects.

\begin{table}[h]
\caption{\label{Tab:combinedsample} Properties of the 
combined spectroscopic sample (including NED entries). Among 403
objects in total, 376 found their matches in the source catalogue of
the VIMOS pre-imaging. We list below their celestial positions
(Columns.~1\&2), spectroscopic redshifts (Column~3), $R$-band
magnitude in the Vega system from the VIMOS pre-imaging
(Column~4), and spectral types (Column~5). The spectral classification
is based on the presence of emission lines in the spectrum; {\tt{0}}
and {\tt{1}} refer to passive and active galaxies, stars are indicated
by {\tt{2}}, and {\tt{99}} indicates bad or missing spectra yielding
no classification.}
\begin{tabular}{ccccc}
\hline\hline ra & dec & z & R & type \\
\hline
  23:12:02.4 & -42:58:58 & 0 & 20.675 & {\tt{2}} \\
  23:12:02.9 & -42:57:22 & 0 & 18.670 & {\tt{2}} \\
  23:12:04.2 & -42:46:22 & 0 & 21.063 & {\tt{2}} \\
  23:12:40.6 & -42:57:27 & 0 & 20.181 & {\tt{2}} \\
  23:12:48.6 & -42:46:03 & 0 & 13.557 & {\tt{2}} \\
  23:12:50.7 & -42:53:48 & 0 & 17.568 & {\tt{2}} \\
  23:12:52.2 & -42:55:29 & 0 & 20.256 & {\tt{2}} \\
  23:12:55.5 & -42:32:02 & 0 & 21.364 & {\tt{2}} \\
  23:12:55.7 & -42:41:38 & 0 & 20.559 & {\tt{2}} \\
  23:13:18.2 & -42:47:20 & 0 & 20.125 & {\tt{2}} \\
  23:13:31.2 & -42:35:55 & 0 & 20.672 & {\tt{2}} \\
  23:13:31.6 & -42:51:50 & 0 & 18.420 & {\tt{2}} \\
  23:13:32.4 & -42:32:50 & 0 & 20.263 & {\tt{2}} \\
  23:13:37.4 & -42:26:55 & 0 & 21.731 & {\tt{2}} \\
  23:13:41.7 & -42:34:23 & 0 & 14.277 & {\tt{2}} \\
  23:13:53.6 & -42:39:34 & 0 & 22.019 & {\tt{2}} \\
  23:13:57.1 & -42:56:05 & 0 & 15.085 & {\tt{2}} \\
  23:14:09.8 & -42:33:26 & 0 & 18.432 & {\tt{2}} \\
  23:14:16.9 & -42:53:21 & 0 & 20.796 & {\tt{2}} \\
  23:14:19.6 & -42:51:26 & 0 & 20.626 & {\tt{2}} \\
  23:14:23.0 & -42:43:08 & 0 & 20.275 & {\tt{2}} \\
  23:14:25.8 & -42:37:18 & 0 & 17.694 & {\tt{2}} \\
  23:14:30.1 & -42:52:05 & 0 & 20.844 & {\tt{2}} \\
  23:14:32.8 & -42:57:58 & 0 & 21.264 & {\tt{2}} \\
  23:14:32.8 & -42:57:58 & 0 & 21.264 & {\tt{2}} \\
  23:14:36.3 & -42:39:18 & 0 & 19.154 & {\tt{2}} \\
  23:14:39.3 & -42:43:26 & 0 & 19.205 & {\tt{2}} \\
  23:14:42.0 & -42:29:14 & 0 & 20.507 & {\tt{2}} \\
  23:14:45.0 & -42:33:41 & 0 & 16.645 & {\tt{2}} \\
  23:14:50.4 & -42:48:57 & 0 & 19.635 & {\tt{2}} \\
  23:15:00.9 & -42:59:45 & 0 & 18.288 & {\tt{2}} \\
  23:15:05.7 & -42:57:27 & 0 & 18.919 & {\tt{2}} \\
  23:15:06.6 & -42:34:21 & 0 & 20.943 & {\tt{2}} \\
  23:15:10.5 & -42:53:24 & 0 & 15.840 & {\tt{2}} \\
  23:15:17.8 & -42:59:11 & 0 & 21.111 & {\tt{2}} \\
  23:15:22.0 & -42:59:25 & 0 & 21.490 & {\tt{2}} \\
  23:15:37.1 & -42:23:47 & 0 & 20.301 & {\tt{2}} \\
  23:15:38.2 & -42:21:50 & 0 & 21.576 & {\tt{2}} \\
  23:15:49.1 & -42:24:15 & 0 & 20.795 & {\tt{2}} \\
  23:15:52.2 & -42:30:56 & 0 & 20.958 & {\tt{2}} \\
  23:15:53.3 & -42:29:01 & 0 & 21.066 & {\tt{2}} \\
  23:14:57.3 & -42:33:30 & 0.0060 & 15.930 & {\tt{1}} \\
  23:15:33.1 & -42:24:40 & 0.0282 & 20.073 & {\tt{0}} \\
  23:15:07.1 & -42:45:59 & 0.0416 & 18.355 & {\tt{1}} \\
  23:13:46.0 & -42:45:22 & 0.0516 & 17.429 & {\tt{0}} \\
\hline\end{tabular}
\end{table}
\addtocounter{table}{-1}
\begin{table}\caption{continued.}
\begin{tabular}{ccccc}
\hline\hline ra & dec & z & R & type \\
\hline
  23:13:48.8 & -42:46:34 & 0.0525 & 15.274 & {\tt{0}} \\
  23:14:04.3 & -42:38:05 & 0.0533 & 16.823 & {\tt{1}} \\
  23:12:15.2 & -42:44:28 & 0.0535 & 17.611 & {\tt{0}} \\
  23:14:11.3 & -42:36:40 & 0.0537 & 20.911 & {\tt{1}} \\
  23:13:48.9 & -42:47:09 & 0.0540 & 15.822 & {\tt{0}} \\
  23:14:33.6 & -42:20:30 & 0.0540 & 17.025 & {\tt{0}} \\
  23:15:52.0 & -42:21:24 & 0.0540 & 16.590 & {\tt{1}} \\
  23:13:34.6 & -42:38:17 & 0.0542 & 18.938 & {\tt{0}} \\
  23:13:41.4 & -42:35:20 & 0.0542 & 19.584 & {\tt{0}} \\
  23:13:25.5 & -42:30:32 & 0.0543 & 16.204 & {\tt{1}} \\
  23:14:23.0 & -42:39:01 & 0.0545 & 17.238 & {\tt{0}} \\
  23:13:57.2 & -42:41:39 & 0.0546 & 15.999 & {\tt{0}} \\
  23:15:30.9 & -42:29:14 & 0.0549 & 16.274 & {\tt{99}} \\
  23:13:13.3 & -42:41:07 & 0.0550 & 15.013 & {\tt{0}} \\
  23:14:16.5 & -42:43:16 & 0.0550 & 17.490 & {\tt{0}} \\
  23:12:03.2 & -42:54:13 & 0.0553 & 15.860 & {\tt{1}} \\
  23:14:25.6 & -42:33:27 & 0.0556 & 17.465 & {\tt{1}} \\
  23:13:49.6 & -42:44:44 & 0.0558 & 17.653 & {\tt{0}} \\
  23:12:53.1 & -42:29:19 & 0.0559 & 18.837 & {\tt{1}} \\
  23:13:02.2 & -42:36:36 & 0.0559 & 15.272 & {\tt{0}} \\
  23:12:40.5 & -42:53:16 & 0.0560 & 16.208 & {\tt{0}} \\
  23:12:41.0 & -42:54:28 & 0.0560 & 18.487 & {\tt{1}} \\
  23:14:00.3 & -42:52:07 & 0.0560 & 18.986 & {\tt{0}} \\
  23:14:26.9 & -42:42:28 & 0.0560 & 16.157 & {\tt{0}} \\
  23:13:13.5 & -42:27:27 & 0.0563 & 16.360 & {\tt{0}} \\
  23:13:58.6 & -42:43:39 & 0.0565 & 14.537 & {\tt{0}} \\
  23:13:16.0 & -42:38:11 & 0.0567 & 16.195 & {\tt{0}} \\
  23:13:54.9 & -42:43:57 & 0.0567 & 18.315 & {\tt{0}} \\
  23:14:59.5 & -42:47:46 & 0.0567 & 19.153 & {\tt{0}} \\
  23:15:00.9 & -42:59:54 & 0.0567 & 17.837 & {\tt{1}} \\
  23:14:25.5 & -42:40:47 & 0.0569 & 15.714 & {\tt{0}} \\
  23:14:03.4 & -42:29:47 & 0.0570 & 17.321 & {\tt{0}} \\
  23:14:21.0 & -42:45:01 & 0.0571 & 17.205 & {\tt{0}} \\
  23:13:06.0 & -42:33:21 & 0.0572 & 18.481 & {\tt{0}} \\
  23:13:37.5 & -42:39:42 & 0.0572 & 20.064 & {\tt{0}} \\
  23:13:50.9 & -42:45:46 & 0.0572 & 16.504 & {\tt{0}} \\
  23:14:04.3 & -42:38:24 & 0.0574 & 17.827 & {\tt{0}} \\
  23:13:53.1 & -42:36:15 & 0.0577 & 15.123 & {\tt{0}} \\
  23:13:54.6 & -42:47:53 & 0.0577 & 19.654 & {\tt{1}} \\
  23:14:03.5 & -42:43:11 & 0.0578 & 17.559 & {\tt{0}} \\
  23:13:29.7 & -42:28:27 & 0.0579 & 15.819 & {\tt{0}} \\
  23:12:19.9 & -42:53:50 & 0.0580 & 15.518 & {\tt{0}} \\
  23:13:52.3 & -42:42:39 & 0.0582 & 19.396 & {\tt{0}} \\
  23:13:58.0 & -42:37:11 & 0.0583 & 19.949 & {\tt{0}} \\
  23:14:36.8 & -42:42:51 & 0.0589 & 17.785 & {\tt{0}} \\
  23:13:38.9 & -43:01:09 & 0.0593 & 17.177 & {\tt{0}} \\
  23:13:54.7 & -42:41:18 & 0.0608 & 17.371 & {\tt{0}} \\
  23:14:49.9 & -42:44:24 & 0.0612 & 19.876 & {\tt{1}} \\
  23:14:01.3 & -42:44:26 & 0.0620 & 16.000 & {\tt{0}} \\
  23:14:00.1 & -42:43:02 & 0.0632 & 18.313 & {\tt{0}} \\
  23:14:11.7 & -42:50:54 & 0.0674 & 20.514 & {\tt{1}} \\
  23:12:55.8 & -42:41:47 & 0.0675 & 19.938 & {\tt{1}} \\
  23:14:21.6 & -42:45:59 & 0.0686 & 15.264 & {\tt{0}} \\
  23:14:43.4 & -42:49:53 & 0.0686 & 18.405 & {\tt{1}} \\
  23:15:09.8 & -42:56:37 & 0.0687 & 17.428 & {\tt{0}} \\
  23:14:18.9 & -42:41:04 & 0.0689 & 18.306 & {\tt{1}} \\
  23:12:04.5 & -42:44:08 & 0.0728 & 19.352 & {\tt{1}} \\
  23:14:12.3 & -42:48:49 & 0.0880 & 17.456 & {\tt{0}} \\
  23:13:36.7 & -43:00:55 & 0.0883 & 19.889 & {\tt{1}} \\
  23:14:39.2 & -42:43:28 & 0.0883 & 18.021 & {\tt{1}} \\
\hline\end{tabular}\end{table}
\addtocounter{table}{-1}
\begin{table}\caption{continued.}
\begin{tabular}{ccccc}
\hline\hline ra & dec & z & R & type \\
\hline
  23:14:16.8 & -42:49:09 & 0.0886 & 17.213 & {\tt{1}} \\
  23:15:06.1 & -42:45:42 & 0.0886 & 20.193 & {\tt{1}} \\
  23:14:28.6 & -42:56:26 & 0.0888 & 17.583 & {\tt{1}} \\
  23:13:29.4 & -42:56:41 & 0.0889 & 20.435 & {\tt{1}} \\
  23:13:29.4 & -42:56:41 & 0.0889 & 20.435 & {\tt{1}} \\
  23:13:12.3 & -42:59:02 & 0.0890 & 19.318 & {\tt{1}} \\
  23:13:15.7 & -42:44:51 & 0.0896 & 16.830 & {\tt{0}} \\
  23:13:13.6 & -42:27:31 & 0.0914 & 16.360 & {\tt{1}} \\
  23:14:50.7 & -42:20:06 & 0.0914 & 18.269 & {\tt{1}} \\
  23:13:25.9 & -42:58:08 & 0.0922 & 18.997 & {\tt{1}} \\
  23:13:27.5 & -42:32:05 & 0.0983 & 17.569 & {\tt{1}} \\
  23:13:57.9 & -42:33:02 & 0.0985 & 19.212 & {\tt{0}} \\
  23:13:57.6 & -42:31:44 & 0.0986 & 16.170 & {\tt{0}} \\
  23:13:41.5 & -42:32:42 & 0.0992 & 21.087 & {\tt{0}} \\
  23:14:03.3 & -42:29:13 & 0.0992 & 16.414 & {\tt{1}} \\
  23:13:41.8 & -42:34:15 & 0.0993 & 15.929 & {\tt{0}} \\
  23:13:49.5 & -42:26:34 & 0.0993 & 20.185 & {\tt{1}} \\
  23:13:55.8 & -42:29:51 & 0.0993 & 19.445 & {\tt{1}} \\
  23:14:33.3 & -42:39:47 & 0.0994 & 20.112 & {\tt{1}} \\
  23:13:46.9 & -42:34:38 & 0.0995 & 15.879 & {\tt{0}} \\
  23:14:13.1 & -42:32:23 & 0.0995 & 18.922 & {\tt{1}} \\
  23:13:43.4 & -42:28:00 & 0.0996 & 17.097 & {\tt{1}} \\
  23:13:55.4 & -42:26:19 & 0.0996 & 19.460 & {\tt{1}} \\
  23:14:09.7 & -42:33:33 & 0.0996 & 17.321 & {\tt{1}} \\
  23:12:54.0 & -42:30:50 & 0.0997 & 19.272 & {\tt{1}} \\
  23:12:51.7 & -42:31:05 & 0.1000 & 18.019 & {\tt{0}} \\
  23:13:33.6 & -42:27:33 & 0.1001 & 19.740 & {\tt{0}} \\
  23:14:07.7 & -42:41:17 & 0.1001 & 17.723 & {\tt{1}} \\
  23:13:04.7 & -42:52:59 & 0.1002 & 17.564 & {\tt{0}} \\
  23:13:12.0 & -42:29:34 & 0.1002 & 15.940 & {\tt{0}} \\
  23:13:14.5 & -42:29:57 & 0.1003 & 17.473 & {\tt{1}} \\
  23:13:52.7 & -42:28:47 & 0.1003 & 16.793 & {\tt{0}} \\
  23:15:22.2 & -42:34:03 & 0.1005 & 18.882 & {\tt{1}} \\
  23:13:52.8 & -42:28:23 & 0.1006 & 16.729 & {\tt{0}} \\
  23:13:54.4 & -42:29:21 & 0.1009 & 16.981 & {\tt{0}} \\
  23:13:14.3 & -42:29:59 & 0.1012 & 17.473 & {\tt{1}} \\
  23:15:52.2 & -42:25:47 & 0.1029 & 17.694 & {\tt{1}} \\
  23:14:25.7 & -42:37:28 & 0.1039 & 18.984 & {\tt{0}} \\
  23:13:56.8 & -42:39:02 & 0.1045 & 17.897 & {\tt{0}} \\
  23:14:28.0 & -42:24:49 & 0.1046 & 20.361 & {\tt{1}} \\
  23:14:22.7 & -42:38:03 & 0.1051 & 16.207 & {\tt{0}} \\
  23:13:50.8 & -42:30:00 & 0.1055 & 19.993 & {\tt{1}} \\
  23:13:46.2 & -42:30:35 & 0.1056 & 18.894 & {\tt{0}} \\
  23:14:20.5 & -42:36:21 & 0.1057 & 18.423 & {\tt{1}} \\
  23:13:52.8 & -42:28:25 & 0.1059 & 16.729 & {\tt{1}} \\
  23:12:17.8 & -42:47:10 & 0.1065 & 18.340 & {\tt{0}} \\
  23:13:48.4 & -42:31:24 & 0.1065 & 20.376 & {\tt{1}} \\
  23:13:48.4 & -42:31:24 & 0.1065 & 20.376 & {\tt{1}} \\
  23:13:52.8 & -42:28:25 & 0.1066 & 16.729 & {\tt{1}} \\
  23:12:13.8 & -42:46:36 & 0.1067 & 16.777 & {\tt{0}} \\
  23:12:46.5 & -42:56:27 & 0.1134 & 18.271 & {\tt{1}} \\
  23:13:57.2 & -42:59:11 & 0.1149 & 19.788 & {\tt{1}} \\
  23:14:14.3 & -42:46:33 & 0.1156 & 16.667 & {\tt{0}} \\
  23:14:59.8 & -42:33:44 & 0.1156 & 19.156 & {\tt{1}} \\
  23:15:06.7 & -42:35:18 & 0.1157 & 19.663 & {\tt{1}} \\
  23:15:11.5 & -42:34:52 & 0.1158 & 17.540 & {\tt{1}} \\
  23:13:41.2 & -42:42:03 & 0.1160 & 19.935 & {\tt{1}} \\
  23:14:28.1 & -42:40:16 & 0.1163 & 17.790 & {\tt{0}} \\
  23:13:39.1 & -42:48:23 & 0.1164 & 18.580 & {\tt{1}} \\
  23:12:41.7 & -42:28:04 & 0.1166 & 20.365 & {\tt{1}} \\
  23:14:30.9 & -42:42:17 & 0.1166 & 16.399 & {\tt{0}} \\
\hline\end{tabular}\end{table}
\addtocounter{table}{-1}
\begin{table}\caption{continued.}
\begin{tabular}{ccccc}
\hline\hline ra & dec & z & R & type \\
\hline
  23:13:24.0 & -42:53:05 & 0.1167 & 18.490 & {\tt{0}} \\
  23:13:19.9 & -42:41:49 & 0.1182 & 21.422 & {\tt{1}} \\
  23:13:23.3 & -42:51:21 & 0.1182 & 17.247 & {\tt{0}} \\
  23:13:07.2 & -42:56:00 & 0.1186 & 18.998 & {\tt{1}} \\
  23:14:23.0 & -42:58:28 & 0.1208 & 17.618 & {\tt{1}} \\
  23:15:08.7 & -42:45:23 & 0.1244 & 20.195 & {\tt{1}} \\
  23:14:16.8 & -42:53:04 & 0.1248 & 17.663 & {\tt{0}} \\
  23:13:34.2 & -42:54:55 & 0.1255 & 19.359 & {\tt{1}} \\
  23:14:42.9 & -42:50:25 & 0.1258 & 17.884 & {\tt{0}} \\
  23:14:27.5 & -42:56:11 & 0.1261 & 20.826 & {\tt{1}} \\
  23:13:31.3 & -42:53:19 & 0.1327 & 18.859 & {\tt{1}} \\
  23:14:21.6 & -42:56:41 & 0.1327 & 20.942 & {\tt{1}} \\
  23:14:25.9 & -42:57:26 & 0.1327 & 18.149 & {\tt{1}} \\
  23:13:55.4 & -42:57:46 & 0.1338 & 21.261 & {\tt{1}} \\
  23:12:55.8 & -42:30:29 & 0.1358 & 20.306 & {\tt{1}} \\
  23:13:10.6 & -42:42:59 & 0.1410 & 20.071 & {\tt{1}} \\
  23:12:55.0 & -42:37:28 & 0.1444 & 18.583 & {\tt{1}} \\
  23:13:52.3 & -42:47:34 & 0.1445 & 20.025 & {\tt{1}} \\
  23:13:50.6 & -42:52:02 & 0.1449 & 17.232 & {\tt{1}} \\
  23:14:28.4 & -42:53:49 & 0.1456 & 20.449 & {\tt{1}} \\
  23:15:00.4 & -42:52:39 & 0.1509 & 17.240 & {\tt{0}} \\
  23:15:06.7 & -42:52:06 & 0.1509 & 19.415 & {\tt{0}} \\
  23:12:02.5 & -42:58:29 & 0.1526 & 17.943 & {\tt{1}} \\
  23:14:25.6 & -42:23:47 & 0.1533 & 19.055 & {\tt{1}} \\
  23:14:56.8 & -42:23:27 & 0.1534 & 18.630 & {\tt{0}} \\
  23:13:50.3 & -42:49:56 & 0.1540 & 17.647 & {\tt{0}} \\
  23:13:49.6 & -42:36:58 & 0.1542 & 18.016 & {\tt{1}} \\
  23:15:37.7 & -42:34:38 & 0.1542 & 18.217 & {\tt{1}} \\
  23:14:30.6 & -42:41:34 & 0.1543 & 18.686 & {\tt{1}} \\
  23:14:37.6 & -42:21:12 & 0.1549 & 19.802 & {\tt{1}} \\
  23:12:43.5 & -42:57:07 & 0.1550 & 17.469 & {\tt{0}} \\
  23:13:09.1 & -42:58:19 & 0.1591 & 18.067 & {\tt{1}} \\
  23:14:48.1 & -42:21:46 & 0.1617 & 19.329 & {\tt{0}} \\
  23:14:20.5 & -42:53:46 & 0.1661 & 19.186 & {\tt{1}} \\
  23:11:45.1 & -42:48:10 & 0.1662 & 19.509 & {\tt{0}} \\
  23:15:40.3 & -42:22:12 & 0.1679 & 20.249 & {\tt{1}} \\
  23:14:21.6 & -42:46:56 & 0.1698 & 18.845 & {\tt{1}} \\
  23:13:11.0 & -42:26:55 & 0.1700 & 20.542 & {\tt{1}} \\
  23:12:39.0 & -42:32:28 & 0.1706 & 18.304 & {\tt{1}} \\
  23:14:01.1 & -42:54:33 & 0.1709 & 19.810 & {\tt{1}} \\
  23:15:53.3 & -42:23:23 & 0.1715 & 18.663 & {\tt{0}} \\
  23:15:34.8 & -43:01:35 & 0.1716 & 18.682 & {\tt{1}} \\
  23:13:52.1 & -42:35:30 & 0.1719 & 19.402 & {\tt{1}} \\
  23:14:06.0 & -42:55:22 & 0.1721 & 20.917 & {\tt{0}} \\
  23:13:46.7 & -42:52:35 & 0.1724 & 18.143 & {\tt{1}} \\
  23:13:47.4 & -42:40:02 & 0.1725 & 20.430 & {\tt{1}} \\
  23:13:57.5 & -42:56:14 & 0.1726 & 18.615 & {\tt{0}} \\
  23:12:55.8 & -42:37:08 & 0.1727 & 20.683 & {\tt{1}} \\
  23:13:28.6 & -42:55:53 & 0.1727 & 19.195 & {\tt{1}} \\
  23:12:48.6 & -42:56:41 & 0.1728 & 19.207 & {\tt{1}} \\
  23:14:07.5 & -43:00:10 & 0.1729 & 19.857 & {\tt{0}} \\
  23:12:49.6 & -42:36:15 & 0.1732 & 19.187 & {\tt{1}} \\
  23:14:08.6 & -43:01:41 & 0.1738 & 19.125 & {\tt{0}} \\
  23:11:48.5 & -42:56:27 & 0.1754 & 18.220 & {\tt{0}} \\
  23:15:00.3 & -42:37:29 & 0.1755 & 18.618 & {\tt{1}} \\
  23:12:06.8 & -42:55:57 & 0.1766 & 19.274 & {\tt{1}} \\
  23:14:55.5 & -42:32:49 & 0.1793 & 19.472 & {\tt{0}} \\
  23:14:56.6 & -42:32:22 & 0.1799 & 19.632 & {\tt{1}} \\
  23:14:36.6 & -42:48:26 & 0.1829 & 18.616 & {\tt{1}} \\
  23:13:31.9 & -42:53:58 & 0.1958 & 19.157 & {\tt{1}} \\
  23:13:58.3 & -42:51:33 & 0.1963 & 18.246 & {\tt{1}} \\
\hline\end{tabular}\end{table}
\addtocounter{table}{-1}
\begin{table}\caption{continued.}
\begin{tabular}{ccccc}
\hline\hline ra & dec & z & R & type \\
\hline
  23:14:09.7 & -42:47:45 & 0.1968 & 19.010 & {\tt{1}} \\
  23:14:26.1 & -42:41:52 & 0.2025 & 17.761 & {\tt{0}} \\
  23:14:39.0 & -42:59:01 & 0.2026 & 19.262 & {\tt{0}} \\
  23:14:28.2 & -42:59:18 & 0.2032 & 18.792 & {\tt{0}} \\
  23:14:40.1 & -43:02:56 & 0.2032 & 20.802 & {\tt{1}} \\
  23:14:13.6 & -42:54:56 & 0.2039 & 19.435 & {\tt{1}} \\
  23:13:52.9 & -43:01:51 & 0.2040 & 19.318 & {\tt{1}} \\
  23:13:09.9 & -42:48:02 & 0.2062 & 19.662 & {\tt{1}} \\
  23:14:32.9 & -42:47:26 & 0.2064 & 18.632 & {\tt{0}} \\
  23:14:33.3 & -42:47:36 & 0.2067 & 19.422 & {\tt{0}} \\
  23:14:32.3 & -42:47:54 & 0.2069 & 19.741 & {\tt{1}} \\
  23:14:14.7 & -42:49:42 & 0.2072 & 18.716 & {\tt{1}} \\
  23:15:10.6 & -42:49:22 & 0.2106 & 18.621 & {\tt{0}} \\
  23:14:39.7 & -42:50:30 & 0.2115 & 18.986 & {\tt{1}} \\
  23:14:59.9 & -42:53:55 & 0.2150 & 19.220 & {\tt{0}} \\
  23:14:42.1 & -42:47:55 & 0.2155 & 17.756 & {\tt{0}} \\
  23:15:00.9 & -42:53:06 & 0.2159 & 18.463 & {\tt{0}} \\
  23:13:48.1 & -42:40:34 & 0.2164 & 19.909 & {\tt{1}} \\
  23:13:54.5 & -42:38:54 & 0.2193 & 18.803 & {\tt{1}} \\
  23:14:06.6 & -42:37:31 & 0.2203 & 19.665 & {\tt{1}} \\
  23:12:19.2 & -42:55:39 & 0.2239 & 20.155 & {\tt{1}} \\
  23:11:47.5 & -42:56:50 & 0.2241 & 19.400 & {\tt{1}} \\
  23:15:31.7 & -43:01:52 & 0.2254 & 18.682 & {\tt{0}} \\
  23:13:52.0 & -42:58:35 & 0.2300 & 20.465 & {\tt{1}} \\
  23:11:54.3 & -42:45:44 & 0.2319 & 17.860 & {\tt{0}} \\
  23:12:43.5 & -42:54:46 & 0.2319 & 18.493 & {\tt{0}} \\
  23:13:27.0 & -42:49:22 & 0.2321 & 18.509 & {\tt{0}} \\
  23:13:24.1 & -43:00:03 & 0.2327 & 19.717 & {\tt{1}} \\
  23:12:50.8 & -42:53:40 & 0.2341 & 19.433 & {\tt{1}} \\
  23:14:50.8 & -43:00:49 & 0.2347 & 21.008 & {\tt{1}} \\
  23:15:26.0 & -42:58:33 & 0.2356 & 19.157 & {\tt{1}} \\
  23:14:28.4 & -43:02:36 & 0.2357 & 20.330 & {\tt{1}} \\
  23:14:35.2 & -43:02:11 & 0.2358 & 20.499 & {\tt{1}} \\
  23:15:34.2 & -42:48:53 & 0.2382 & 20.890 & {\tt{1}} \\
  23:12:05.2 & -42:53:59 & 0.2452 & 19.695 & {\tt{0}} \\
  23:13:15.0 & -42:42:33 & 0.2471 & 19.196 & {\tt{1}} \\
  23:11:53.9 & -42:55:00 & 0.2481 & 20.513 & {\tt{1}} \\
  23:15:00.3 & -42:42:17 & 0.2519 & 20.353 & {\tt{1}} \\
  23:14:43.1 & -43:02:29 & 0.2520 & 20.199 & {\tt{1}} \\
  23:15:02.9 & -42:36:25 & 0.2539 & 18.991 & {\tt{1}} \\
  23:15:32.0 & -43:02:24 & 0.2546 & 19.683 & {\tt{1}} \\
  23:12:17.7 & -42:49:00 & 0.2571 & 20.310 & {\tt{1}} \\
  23:14:26.0 & -42:31:31 & 0.2598 & 15.860 & {\tt{1}} \\
  23:14:57.9 & -42:35:55 & 0.2598 & 18.825 & {\tt{0}} \\
  23:15:00.2 & -42:37:18 & 0.2603 & 20.911 & {\tt{1}} \\
  23:14:34.9 & -42:25:11 & 0.2605 & 20.291 & {\tt{1}} \\
  23:13:56.5 & -42:32:16 & 0.2612 & 21.443 & {\tt{1}} \\
  23:14:57.3 & -42:38:58 & 0.2613 & 19.478 & {\tt{1}} \\
  23:13:54.6 & -42:27:37 & 0.2614 & 20.947 & {\tt{1}} \\
  23:13:10.4 & -42:57:57 & 0.2620 & 19.264 & {\tt{0}} \\
  23:13:23.5 & -42:50:47 & 0.2630 & 20.389 & {\tt{1}} \\
  23:14:52.2 & -42:51:42 & 0.2687 & 19.410 & {\tt{0}} \\
  23:14:31.7 & -43:01:20 & 0.2742 & 19.791 & {\tt{1}} \\
  23:13:59.7 & -42:33:24 & 0.2776 & 19.795 & {\tt{1}} \\
  23:12:05.2 & -42:45:15 & 0.2879 & 19.027 & {\tt{0}} \\
  23:13:13.7 & -42:45:49 & 0.2879 & 20.797 & {\tt{1}} \\
  23:13:03.9 & -42:59:26 & 0.2881 & 20.503 & {\tt{0}} \\
  23:13:23.2 & -42:43:41 & 0.2899 & 20.717 & {\tt{0}} \\
  23:14:45.7 & -42:45:03 & 0.2904 & 21.152 & {\tt{0}} \\
  23:13:27.1 & -42:29:05 & 0.2964 & 20.484 & {\tt{0}} \\
  23:15:18.9 & -42:50:48 & 0.3076 & 20.854 & {\tt{1}} \\
\hline\end{tabular}\end{table}
\addtocounter{table}{-1}
\begin{table}\caption{continued.}
\begin{tabular}{ccccc}
\hline\hline ra & dec & z & R & type \\
\hline
  23:15:36.5 & -42:32:07 & 0.3096 & 20.141 & {\tt{1}} \\
  23:12:54.6 & -42:35:53 & 0.3097 & 17.726 & {\tt{1}} \\
  23:15:44.1 & -42:31:40 & 0.3098 & 20.391 & {\tt{1}} \\
  23:14:25.0 & -43:00:24 & 0.3099 & 20.259 & {\tt{1}} \\
  23:13:13.8 & -42:50:36 & 0.3110 & 20.712 & {\tt{1}} \\
  23:13:26.8 & -42:37:56 & 0.3114 & 20.390 & {\tt{1}} \\
  23:12:59.4 & -42:38:40 & 0.3124 & 20.459 & {\tt{1}} \\
  23:15:25.1 & -42:22:34 & 0.3209 & 19.837 & {\tt{1}} \\
  23:14:19.3 & -42:52:42 & 0.3211 & 19.695 & {\tt{1}} \\
  23:14:57.4 & -42:20:34 & 0.3218 & 21.297 & {\tt{1}} \\
  23:15:21.9 & -42:29:57 & 0.3261 & 19.562 & {\tt{0}} \\
  23:14:26.6 & -42:30:10 & 0.3267 & 20.302 & {\tt{1}} \\
  23:11:59.6 & -42:49:34 & 0.3311 & 19.380 & {\tt{1}} \\
  23:14:27.2 & -42:35:53 & 0.3431 & 20.116 & {\tt{0}} \\
  23:14:33.4 & -42:22:02 & 0.3434 & 21.012 & {\tt{0}} \\
  23:14:33.4 & -42:22:02 & 0.3434 & 21.012 & {\tt{0}} \\
  23:14:47.3 & -42:35:36 & 0.3444 & 20.094 & {\tt{1}} \\
  23:14:36.3 & -42:44:35 & 0.3477 & 20.238 & {\tt{0}} \\
  23:15:16.8 & -42:50:20 & 0.3477 & 20.209 & {\tt{1}} \\
  23:14:36.3 & -42:44:38 & 0.3483 & 20.242 & {\tt{0}} \\
  23:14:53.4 & -43:02:11 & 0.3485 & 19.508 & {\tt{1}} \\
  23:13:46.5 & -42:40:16 & 0.3496 & 18.877 & {\tt{1}} \\
  23:13:31.9 & -42:54:05 & 0.3498 & 18.929 & {\tt{0}} \\
  23:12:58.7 & -42:39:42 & 0.3525 & 20.191 & {\tt{0}} \\
  23:14:26.2 & -42:31:06 & 0.3529 & 19.051 & {\tt{0}} \\
  23:13:06.6 & -42:44:03 & 0.3531 & 20.252 & {\tt{1}} \\
  23:14:15.5 & -43:00:53 & 0.3540 & 20.507 & {\tt{0}} \\
  23:14:48.4 & -42:46:51 & 0.3546 & 19.719 & {\tt{1}} \\
  23:14:09.8 & -42:49:59 & 0.3550 & 20.905 & {\tt{1}} \\
  23:11:58.8 & -42:47:46 & 0.3603 & 19.840 & {\tt{1}} \\
  23:12:59.3 & -42:44:10 & 0.3633 & 21.117 & {\tt{1}} \\
  23:14:26.0 & -42:31:27 & 0.3681 & 15.860 & {\tt{1}} \\
  23:14:35.1 & -42:34:24 & 0.3681 & 14.351 & {\tt{1}} \\
  23:13:14.0 & -42:45:09 & 0.3693 & 20.695 & {\tt{1}} \\
  23:15:17.8 & -42:59:07 & 0.3693 & 21.111 & {\tt{1}} \\
  23:13:09.9 & -42:45:22 & 0.3699 & 20.293 & {\tt{0}} \\
  23:14:54.4 & -42:46:34 & 0.3700 & 19.926 & {\tt{1}} \\
  23:15:05.3 & -42:41:33 & 0.3703 & 19.499 & {\tt{1}} \\
  23:14:25.4 & -42:23:41 & 0.3709 & 17.718 & {\tt{1}} \\
  23:15:56.2 & -42:30:13 & 0.3715 & 21.794 & {\tt{1}} \\
  23:13:02.4 & -42:39:02 & 0.3836 & 19.714 & {\tt{1}} \\
  23:13:06.4 & -42:42:18 & 0.3839 & 20.450 & {\tt{1}} \\
  23:15:34.8 & -42:22:52 & 0.3885 & 21.098 & {\tt{1}} \\
  23:15:47.0 & -42:29:20 & 0.3887 & 20.422 & {\tt{0}} \\
  23:15:28.0 & -42:25:10 & 0.3897 & 19.648 & {\tt{0}} \\
  23:15:18.6 & -42:58:05 & 0.3939 & 21.943 & {\tt{1}} \\
  23:15:14.6 & -42:48:29 & 0.4059 & 19.947 & {\tt{1}} \\
  23:15:14.6 & -42:48:25 & 0.4060 & 19.947 & {\tt{1}} \\
  23:15:16.5 & -43:03:05 & 0.4074 & 21.331 & {\tt{0}} \\
  23:15:32.9 & -42:31:21 & 0.4103 & 20.725 & {\tt{1}} \\
  23:14:30.0 & -42:29:01 & 0.4109 & 19.580 & {\tt{1}} \\
  23:14:52.9 & -42:38:05 & 0.4140 & 21.190 & {\tt{1}} \\
  23:15:58.0 & -42:33:05 & 0.4140 & 19.613 & {\tt{0}} \\
  23:14:32.7 & -42:58:01 & 0.4154 & 20.151 & {\tt{1}} \\
  23:15:08.2 & -42:57:07 & 0.4156 & 20.690 & {\tt{1}} \\
  23:15:29.7 & -43:00:20 & 0.4200 & 20.058 & {\tt{1}} \\
  23:14:50.9 & -42:50:39 & 0.4202 & 20.784 & {\tt{1}} \\
  23:14:43.4 & -42:59:57 & 0.4213 & 20.840 & {\tt{0}} \\
  23:14:56.0 & -42:47:28 & 0.4213 & 20.795 & {\tt{1}} \\
  23:14:18.4 & -43:01:41 & 0.4267 & 20.660 & {\tt{1}} \\
  23:12:42.3 & -42:40:27 & 0.4387 & 21.078 & {\tt{0}} \\
\hline\end{tabular}\end{table}
\addtocounter{table}{-1}
\begin{table}\caption{continued.}
\begin{tabular}{ccccc}
\hline\hline ra & dec & z & R & type \\
\hline
  23:13:23.4 & -42:48:42 & 0.4402 & 21.961 & {\tt{1}} \\
  23:14:20.8 & -42:59:45 & 0.4428 & 19.950 & {\tt{0}} \\
  23:13:40.1 & -42:50:28 & 0.4448 & 20.298 & {\tt{0}} \\
  23:15:34.8 & -42:33:30 & 0.4457 & 20.220 & {\tt{1}} \\
  23:14:52.0 & -42:25:43 & 0.4497 & 21.585 & {\tt{1}} \\
  23:14:47.7 & -42:30:43 & 0.4590 & 20.990 & {\tt{1}} \\
  23:14:08.4 & -42:35:55 & 0.4630 & 21.650 & {\tt{1}} \\
  23:12:10.0 & -42:50:09 & 0.4655 & 20.694 & {\tt{1}} \\
  23:14:40.1 & -42:43:56 & 0.4725 & 21.152 & {\tt{1}} \\
  23:15:02.8 & -42:35:23 & 0.4769 & 20.301 & {\tt{0}} \\
  23:14:40.4 & -42:53:12 & 0.4779 & 20.478 & {\tt{1}} \\
  23:14:47.5 & -42:22:50 & 0.4790 & 20.507 & {\tt{1}} \\
  23:15:23.6 & -42:35:20 & 0.4799 & 20.294 & {\tt{1}} \\
  23:14:43.9 & -42:38:17 & 0.4800 & 20.558 & {\tt{0}} \\
  23:15:38.6 & -42:35:08 & 0.4859 & 21.017 & {\tt{1}} \\
  23:12:16.3 & -42:59:35 & 0.4909 & 20.797 & {\tt{1}} \\
  23:13:48.6 & -42:30:57 & 0.5318 & 20.682 & {\tt{0}} \\
  23:13:24.6 & -42:37:28 & 0.5337 & 21.521 & {\tt{1}} \\
  23:14:18.4 & -43:01:44 & 0.5363 & 20.660 & {\tt{1}} \\
  23:12:05.4 & -42:57:50 & 0.5540 & 20.619 & {\tt{1}} \\
  23:15:00.8 & -42:51:43 & 0.5631 & 21.102 & {\tt{0}} \\
  23:13:48.0 & -42:56:55 & 0.5735 & 20.718 & {\tt{1}} \\
  23:12:58.1 & -42:28:48 & 0.5788 & 20.618 & {\tt{1}} \\
  23:14:40.2 & -42:29:44 & 0.6130 & 21.816 & {\tt{0}} \\
  23:14:00.9 & -42:46:01 & 0.6278 & 21.765 & {\tt{1}} \\
  23:13:16.2 & -42:43:26 & 0.6715 & 20.965 & {\tt{1}} \\
  23:13:35.8 & -42:46:29 & 3.4000 & 20.961 & {\tt{1}} \\
\hline\end{tabular}\end{table}

\end{appendix}
\end{document}